\documentclass{jfm}
\usepackage{graphicx}
\usepackage{epstopdf, epsfig}
\usepackage{upgreek} 
\usepackage{lineno}
\usepackage{color}

\newcommand\putfig[2]{\begin{tabular}[t]{@{}l@{}}#1\\#2\end{tabular}}
\newcommand{\solidrule}[1][1cm]{\rule[0.5ex]{#1}{.8pt}}
\newcommand{\figline}{\protect \solidrule[6mm] \hspace{0.7mm}}
\newcommand{\figldash}{\protect{\solidrule[1.5mm]\hspace{1.mm}\solidrule[1.5mm]\hspace{1.mm}\solidrule[1.5mm]\hspace{0.7mm}}}
\newcommand{\figsdash}{\protect {\solidrule[0.5mm]\hspace{0.7mm}\solidrule[0.5mm]\hspace{0.7mm}\solidrule[0.5mm]\hspace{0.7mm}\solidrule[0.5mm]\hspace{0.7mm}\solidrule[0.5mm]\hspace{0.7mm}}}
\newcommand{\figddash}{\protect {\solidrule[1.2mm]\hspace{0.7mm}\solidrule[0.5mm]\hspace{0.7mm}\solidrule[1.2mm]\hspace{0.7mm}\solidrule[0.5mm]\hspace{0.7mm}}}

\shorttitle{Modulation of turbulence regeneration cycle by finite-size particles}
\shortauthor{G. Wang, M. Abbas, and E. Climent}

\title{Modulation of the regeneration cycle by neutrally buoyant finite-size particles}
\author{G. Wang\aff{1,2,3},
  M. Abbas\aff{2,3}
  \corresp{\email{Micheline.Abbas@ensiacet.fr}}  
 \and E. Climent\aff{1,3}
 }

\affiliation{\aff{1} Institut de M\'{e}canique des Fluides de Toulouse (IMFT) - Universit\'{e} de Toulouse, CNRS-INPT-UPS, Toulouse FRANCE
\aff{2} Laboratoire de G\'{e}nie Chimique - Universit\'{e} de Toulouse, CNRS-INPT-UPS, Toulouse FRANCE
\aff{3} FERMaT, Universit\'e de Toulouse, CNRS, INPT, INSA, UPS, Toulouse, France}

\begin{document}

\maketitle

\begin{abstract}

Direct numerical simulations of turbulent suspension flows are carried out with the Force-Coupling Method in plane Couette and pressure-driven channel configurations. Dilute to moderately concentrated suspensions of neutrally buoyant finite-size ($L_y/d=20$) spherical particles are considered when the Reynolds number is slightly above the laminar-turbulent transition. Tests performed with synthetic streaks, in both turbulent channel and Couette flows, show clearly that particles trigger the instability in channel flow whereas the plane Couette flow becomes laminar. Moreover, we have shown that particles have a pronounced impact on pressure-driven flow through a detailed temporal and spatial analysis whereas they have no significant impact on plane Couette flow configuration. The substantial difference between both flows is related to spatial preferential distribution of particles in the large scale rolls ($inactive$ motion) in Couette flow, whereas they are accumulated in the ejection ($active$ motion) regions in pressure-driven flow. Through investigation of particle modification on two distinct flow configurations, we were able to show the specific response of turbulent structures and the modulation of the fundamental mechanisms composing the regeneration cycle in the buffer layer of near-wall turbulence. Especially for pressure-driven flow, the particles enhance the lift-up and let it act continuously whereas the particles do no significantly alter the streak breakdown process. The reinforcement of the streamwise vortices is attributed to the vorticity stretching term by the wavy streaks. The smaller and more numerous wavy streaks enhance the vorticity stretching and consequently strengthen the circulation of large scale streamwise vortex in suspension flow.

\end{abstract}


\section{Introduction}


The experiments of \citet{matas2003transition} have shed the light on the non-monotonous effect of particles on laminar-turbulent flow transition, depending on the particle-to-pipe size ratio and on suspension volumetric concentration. A small amount of neutrally buoyant finite-size particles allowed sustaining the turbulent state and decreasing the transition threshold significantly. Almost a decade later, particle-resolved numerical simulations provided some evidences that at moderate concentration, particles have a significant impact on the unsteady nature of the flow, enhancing the transverse turbulent stress components and modifying the flow rotational structures \citep{loisel2013effect, yu2013numerical, lashgari2015transition}. The effect of particles on the transition of Couette flow is not yet well characterized. Recent experiments from \citet{majji2016flow} have shown that particles do not have a significant impact on the transition path in Taylor-Couette flow, if the particle concentration is low and the particle size is relatively small compared to the Couette gap. With larger particles (8 times smaller than the Couette gap), \citet{linares2017effects} have shown that particles do not change the transition threshold of a cylindrical turbulent Couette flow at $10\%$ volumetric concentration. Consistent with this finding, flow statistics performed on moderately concentrated turbulent plane Couette flow (slightly above the transition threshold), have revealed that there is no significant difference between single- and two-phase flows at equivalent effective Reynolds number \citep{wang2017modulation}.\\

If the size of the particles is large enough compared to the size of energetic eddies in a turbulent flow, the local flow streamlines are significantly modified (as would not be the case for pointwise particles). The rigid body constraints from finite-size particles influence the turbulent kinetic energy budget in two competing ways: they add perturbations that increase shear production of turbulence and simultaneously increase viscous dissipation \citep{qureshi2007turbulent, bellani2012shape}. The perturbations induced by the particles depend on their locations: their magnitude increases with the local flow strain rate. The spatial distribution of neutrally buoyant particles depends on the flow configuration (turbulent Couette or channel flow). Indeed, in addition to the turbulent dispersion that particles undergo, they are experiencing a lift force due to finite flow inertia at the particle scale. This lift force is normal to the walls, and its orientation depends itself on the flow configuration. Therefore particles are preferentially located either in the $active$ region, i.e. near the walls in pressure-driven channel flow \citep{loisel2013effect}, or in the $inactive$ region, i.e. away from the walls in Couette flow \citep{wang2017modulation}. \\


Even though the stability in Couette and channel single-phase flows is different, they share at high Reynolds numbers some common turbulence features in the near wall regions. \textcolor{black}{In the inner region of a boundary layer, the turbulent motion are consisted of an $active$ part and an $inactive$ part based on \citet{townsend1980structure}.} Near the walls, the $active$ motion contains eddies with streamwise characteristic length of the order of $1000$ wall units in highly turbulent flows that constitute the essential contribution to the Reynolds shear stress ($-\overline{u^\prime v^\prime}$). Statistical properties of the $active$ motion are universal functions of the friction velocity $u_\uptau$ whereas the $inactive$ motion gives no correlation between $u^\prime$ and $v^\prime$ and it is mainly related to the flow geometry \citep{bradshaw1967inactive, jimenez2011cascades, panton2001overview, tuerke2013simulations}. The essential difference in both flow configurations is due to the mean velocity ($\overline{u}$) profile which is anti-symmetric (resp. symmetric) in plane Couette (resp. pressure-driven) flow with respect to the midplane. The production term ($-\overline{u^\prime v^\prime} d\overline{u}/dy$) in the turbulent kinetic energy equation has different roles according to the flow configuration. In Couette flow, the fluid is pumped away from one wall to the other one, extracting energy from the mean flow, leading to the enhancement of turbulent structures \citep{papavassiliou1997interpretation}. However in pressure-driven channel flow, the shear layers are divided into two regions and the production is of opposite sign in both channel halves, making the turbulent structures relatively independent on each wall. \\

\begin{figure}
  \centerline{\includegraphics[trim = 0mm 0mm 0mm 0mm,width=10.0cm]{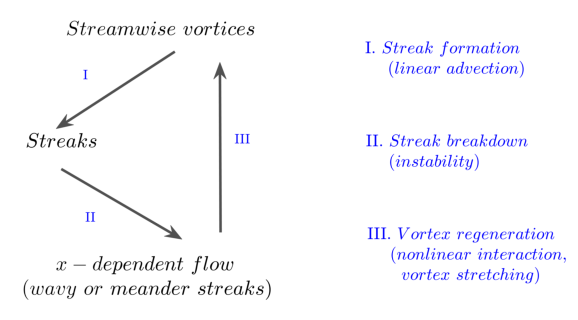}}
  \caption{Sketch of the regeneration cycle sub-steps}
\label{fig:Figure_1}
\end{figure}

The temporal and spatial development of wall turbulence consists of a self-sustained process, namely the near-wall regeneration cycle (located in the vicinity of the non-slip boundary condition $20<y^+<60$, see \citet[]{waleffe1997self}). During this complete cycle, coherent large-scale streaks and alternating staggered rotating vortices sustain each other, altogether having impact on the wall friction. This cycle has been demonstrated to be independent of the outer layer: it can survive without any input from the core flow. Indeed \citet{jimenez1999autonomous} carried out some simulations after removing all the fluctuations from the velocity field above $y^+=60$  (in a channel flow), and after hundreds of time units, they observed an almost unchanged turbulent flow compared to the original one. The regeneration cycle consists of three sequential sub-processes sketched in figure \ref{fig:Figure_1}: streak formation, streak breakdown and streamwise vortex regeneration. The streaks are generated by a linear process, the so-called lift-up effect, whereas the following two processes are the result of non-linear interactions. \\

In order to understand how particles affect the flow turbulence and the transition from one regime to another, we are concerned in this paper with their impact on the regeneration cycle. \citet{klinkenberg2013numerical} have shown that inertial pointwise particles modify the transition to turbulence not by altering the lift-up effect but rather by modifying the dynamics of the oblique waves necessary for the streaks regeneration and breakdown. In this work, we consider the effect of neutrally buoyant finite-size particles on the regeneration cycle, in turbulent flows slightly above the transition limit of single-phase flows (Reynolds number equal to 500 for Couette and 2600 for channel flows). Numerical simulations are performed in a domain (so-called miniunit) which contains one set of coherent structures sufficient to sustain the flow turbulence. The size of this miniunit is different for both flows and it follows the findings of \citet{jimenez1991minimal} and \citet{hamilton1995regeneration}. The coupling between the fluid motion and the particle dynamics is taken into account using the Force Coupling Method \citep{climent2009force}. Neutrally buoyant particles $20$ times smaller than the Couette gap or channel height are considered here at moderate volumetric concentration from $1\%$ up to $10\%$.\\

The paper is organized as follows. Section \ref{sec:Methodology} summarizes the numerical configurations in both single- and two-phase flows. In section \ref{sec:transition}, we show how particles affect the laminar-turbulent transition by using specific initial conditions for each flow configuration. Then, we discuss the effect of particles on the flow energy modulation in section \ref{sec:energy} and on the different stages of the regeneration cycle in section \ref{sec:regeneration_cycle}. Both analogies and divergences between Couette and pressure-driven flows are discussed all along the paper before conclusion.

\section{Suspension flow configurations}\label{sec:Methodology}

The coupling of fluid flow and particle dynamics follows the Force Coupling Method (FCM), as described in \citet{wang2017modulation}. The validation tests were carefully detailed. The method is valid to study suspension Couette or pressure-driven channel flows, with particle Reynolds numbers ($\Rey_p$ is defined in Table \ref{tab:Table_1}) up to $10$ and particle volume fraction less than $20\%$. \\

Couette flow was driven by two walls moving at equal and opposite velocities. Pressure-driven channel flow was generated by imposing a global pressure drop in the streamwise direction, that is timely tuned in order to maintain constant flow rate. In both flow configurations, $x$ and $z$ are respectively the streamwise and spanwise flow directions, with periodic boundary conditions (the so-called homogeneous directions) while $y$ stands for the wall-normal or velocity gradient direction. Turbulent flow simulations were performed using a so-called ``miniunit" configuration, which is the minimal geometric domain that is sufficient to accommodate the self-sustained flow structures for single-phase turbulence, while allowing reasonable time for the computation of suspension flows with finite-size particles. This minimal simulation domain was carefully examined in Couette flow configuration by \citet{hamilton1995regeneration} and pressure-driven flow by \citet{jimenez1991minimal}. In both cases, the spanwise length is larger than 100 wall units which corresponds to the spanwise characteristic spacing between two coherent structures. The length and velocity in wall units are $y^+ \equiv y u_{\uptau}/{\nu}$, and $u^+ \equiv {u}/u_{\uptau}$, where $u_{\uptau}=\sqrt{{\uptau}_w/{\rho}}$ is the friction velocity based on the wall shear stress and fluid density.  \\

\textcolor{black}{Through this work, we use Fourier decomposition to study the regeneration cycle. We performed modal analysis of the flow fluctuating energy. The Fourier decomposition of the energy, as introduced by \citet{hamilton1995regeneration} over two periodic directions, streamwise and spanwise, is written as follows:}

\begin{equation}\label{eq:fft_2d} 
M(k_x=m\alpha ,k_z=n\beta) \equiv  \{ \int ^{Y_2}_{Y_1} [ \widehat{u^\prime}^2(m\alpha,y,n\beta) + \widehat{v^\prime}^2(m\alpha,y,n\beta)+\widehat{w^\prime}^2(m\alpha,y,n\beta)]dy \} ^{1/2}
\end{equation}

\textcolor{black}{where $Y_1$ and $Y_2$ stand for the integration bounds in wall-normal direction.$(\alpha,\beta)$ are the fundamental wavenumbers in streamwise and spanwise directions defined as $(2\pi/L_x,2\pi/L_z)$, and $m$ and $n$ are integers. Any turbulent structure can be represented by one mode $(m\alpha,n\beta)$. For instance, the mode $(0,n\beta)$ with $n\neq0$ is an x-independent structure and the mode $(m\alpha,n\beta)$ with $m \neq 0$ is the x-dependent structure (e.g. streaks confined in the streamwise direction).}\\

Table \ref{tab:Table_1} contains a summary of all the parameters selected for this study. Through all the paper, we note $C$ for plane Couette and $P$ for pressure-driven channel flows. The size ratio between the Couette gap or channel height and the particle diameter is $L_y/d=20$ in most cases. Particles experience turbulent fluctuations, and their inertia can be characterized by the dimensionless Stokes number $St_{turb}=\uptau_p^+/\uptau_f^+$, where $\uptau_p^+$ is the particle relaxation time scale in response to the turbulent flow forcing which characteristic time scale is $\uptau_f^+$. The latter is considered here as the ratio between the characteristic size of the large scale streamwise vortices $L_y^+$ in Couette or $L_y^+/2$ in channel flow, and the characteristic velocity fluctuation scale $max(v^{\prime +} , w^{\prime +})$ in the flow cross-section. The ability of the FCM to capture accurately the particle response to flow fluctuations was tested in \citet{wang2017modulation}, where the motion of a rigid particle submitted to an external oscillating force \textcolor{black}{($\mathbf{F}_{ext}(t)=6 \pi \mu a \mathbf{u}_0 \sin (\omega t)$, where $\mathbf{u}_0$ is a constant vector)} in a still fluid was considered. The numerical solution of the particle motion was in a good agreement with the theoretical prediction when the ratio of the particle radius to the developed Stokes layer thickness $\delta^2 \equiv \omega a^2/\nu$ was less than $2$. This ratio is directly related to the particle Reynolds and Stokes numbers, i.e. $\delta^2 = 9 \pi St_{turb}/(\rho_p/\rho_f+1/2)$. In the simulations considered for this work, with neutrally buoyant particles, $\delta^2$ is always below $0.5$ in Couette and $1.4$ in pressure-driven flows. \textcolor{black}{The Reynolds number used in both configurations are comparable, which is $40\%$ and $50\%$ higher than its transitional Reynolds number (estimated from figure \ref{fig:Figure_2}).}
\\

\begin{table}
  \begin{center}
\def~{\hphantom{0}}
\begin{tabular}{cccccccccccc}
\multicolumn{12}{c}{\begin{tabular}[c]{@{}c@{}} $Couette(C)$ \\ $L_y/d=20$, $L_x \times L_y \times L_z=2.85 \times 1.0 \times 1.91$,$N_x \times N_y \times N_z=382 \times 134 \times 256$\end{tabular}}      \\
\multicolumn{12}{c}{\begin{tabular}[c]{@{}c@{}}$pressure-driven(P)$\\ $L_y/d=20$, $L_x \times L_y \times L_z=1.57 \times 1.0 \times 0.63$,$N_x \times N_y \times N_z=158 \times 106 \times 64$\end{tabular}} \\
\\
$~Case~$             & $~N_p~$         & $~\overline{\Phi}(\%)~$         & $U_{bulk}$         & $~u_{\uptau}~$         & $~L_y^+~$   		 & $~d^+~$      & $~\Rey_b~$        & $\Rey_{\uptau}$        & $\Rey_{p(max)}$        & $St_{(max)}$   & $St_{turb}$     \\
\\
$C500-0~$           & 0             & 0                             & 0.5                & 0.040              & 80          & --                & 500           & 40                 & --                 & --               & --\\
$C500-5~$           & 3968          & 5                             & 0.5                & 0.041              & 82          & 4.1              & 500           & 41                 & 4.4                & 0.97             & 0.017\\
$C500-10$          & 7936          & 10                            & 0.5                & 0.042              & 84          & 4.2              & 500           & 42                 & 4.4                & 0.97             & 0.017\\
\\
$P2600-0$          & 0             & 0                             & 0.5                & 0.048              & 187         & --             & 2600          & 94                 & --                 & --               & --\\
$P2600-1$          & 151           & 1                             & 0.5                & 0.052              & 203         & 10.15            & 2600          & 102                & 10.6               & 2.36              & 0.068\\
$P2600-5$          & 757           & 5                             & 0.5                & 0.056              & 218         & 10.9~             & 2600          & 109                & 7.8                & 1.73              & 0.074
\end{tabular}
  \caption{Parameters of the numerical simulations. The Reynolds number $\Rey_b\equiv U_{bulk}h/\nu$ for Couette flow and $\Rey_b \equiv Q/\nu$ for channel flow. $h=L_y/2$ is half of the Couette gap or channel height. In channel flow, the flow rate per unit depth is $Q=4U_{bulk}L_y/3$. $U_{bulk}$ is the velocity of the moving walls in Couette flow whereas it is half the central velocity that the channel flow would have if the flow was laminar. The Reynolds number based on the friction velocity and on the channel half-width is $\Rey_\tau\equiv u_{\uptau}h/\nu$. The particle Reynolds number $\Rey_p \equiv \Gamma a^2/\nu$ based on local shear rate $\Gamma=\vert {d\overline{u}}/{dy} \vert$, and the Stokes number $St \equiv 2\rho_p/(9\rho_f)\Rey_p$ are low near the Couette and channel centers and they are maximum near the walls where the shear rate is the highest. The maximum particle Reynolds and Stokes numbers are based on the shear rate calculated at one particle diameter away from the walls. }
  \label{tab:Table_1}
  \end{center}
\end{table}

Table \ref{tab:Table_1} contains also the Stokes number based on the local shear which achieves its maximum value near the channel or Couette walls, where the shear rate is the highest. It is nearly 1 in Couette and 2.4 in channel flow. Note that the results of simulations in Couette flow with $1 < St < 4 (4< d^+<8)$ reported in \citet{wang2017modulation} were similar to those obtained in the present paper at $St \approx 1 (d^+ \approx 4)$.

\section{Particle effect on the transition}\label{sec:transition}

\begin{figure}
  \centerline{\includegraphics[trim = 0mm 0mm 0mm 0mm,width=13.0cm]{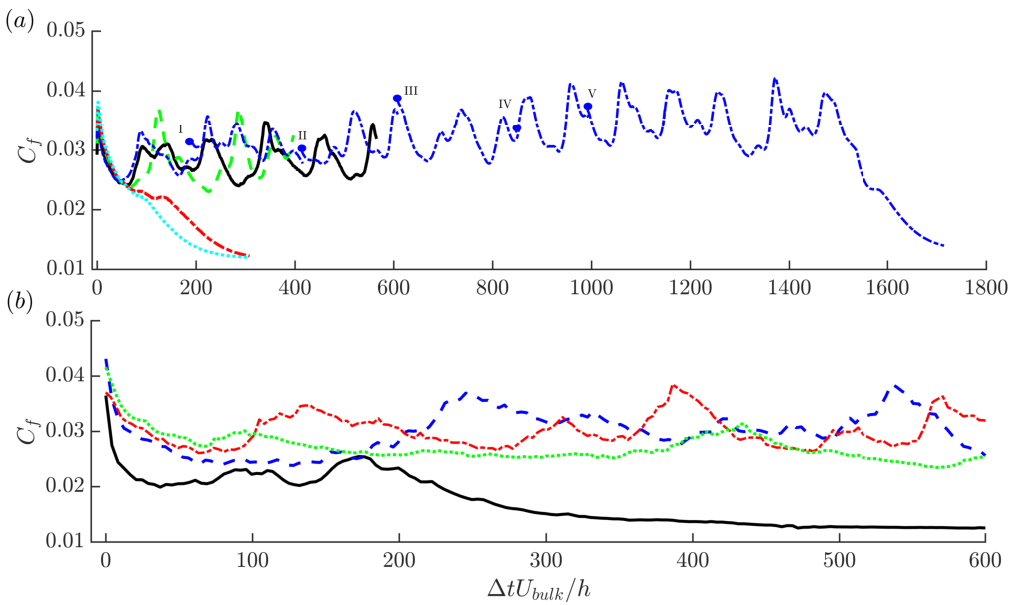}}
  \caption{Effect of neutrally buoyant particles on the laminar-turbulent transition threshold, as depicted from the temporal evolution of $C_f$, after decreasing $\Rey_b$ in $(a)$ Couette flow and $(b)$ pressure-driven flow. The initial flow configuration of the Couette (resp. channel) flow is taken from a fully-turbulent simulation at $\Rey_b=500$ (resp. 2300). $(a)$: \figline $\Rey_b=500$, $\mathit{\Phi}=5\%$; \textcolor{green}{\figldash} $\Rey_b=470$, $\mathit{\Phi}=10\%$; \textcolor{red}{\figddash} $\Rey_b=455$, $\mathit{\Phi}=10\%$; \textcolor{cyan}{\figsdash} $\Rey_b=440$, $\mathit{\Phi}=10\%$; \textcolor{blue}{\figddash} $\Rey_b=455$ to $345$, $\mathit{\Phi}=5\%$ and $I$ to $V$ corresponding to $\Rey_b=455$, $415$, $390$, $365$ and $355$. $(b)$: \figline $\Rey_b=2000$, $\mathit{\Phi}=1\%$; \textcolor{blue}{\figldash} $\Rey_b=2000$, $\mathit{\Phi}=5\%$; \textcolor{red}{\figddash} $\Rey_b=1700$, $\mathit{\Phi}=5\%$; \textcolor{green}{\figsdash} $\Rey_b=1500$, $\mathit{\Phi}=5\%$.}
\label{fig:Figure_2}
\end{figure}

A theoretical analysis of flow stability in the presence of freely moving finite-size particles is actually impossible, from a mathematical point of view. For this reason we determined the transition threshold from an engineering point of view, by considering a fully-developed turbulent flow experiencing successive reductions of the Reynolds number down to a limit where the flow becomes eventually laminar. Every time the Reynolds number was decreased, the simulation was run for longer than $500$ time units. Transition of single-phase flow was observed at $\Rey_{c\_C}\sim 320$ for Couette and at $\Rey_{c\_P}\sim 2200$ for pressure-driven flows. It should be kept in mind that, first the value of the critical Reynolds number depends on the simulation domain because periodic boundary conditions influence interactions between large scale vortices. Second, the relaminarization is a process that may occur randomly. Therefore the determination of a ``rigourous" laminar-turbulent transition threshold (which is not the main scope of the present paper) would require a large amount of simulations to form statistics. Instead, an indicative threshold is determined in order to assess the impact of the particle presence on the flow features, and to evaluate qualitatively transition delay or not. \\

In the range of investigated suspension flow parameters (particle size and volumetric concentration), particles are expected to decrease significantly the laminar-turbulent transition threshold in pressure-driven flow based on the experiments of \citet{matas2003transition}, in opposite to Couette flow where particles seem to not affect the flow stability (see our previous study of \citet{wang2017modulation}). The wall friction coefficient $C_f$ (summed on both walls) was considered as an indicator of current flow regime, $C_f=2\overline{\uptau}_w/(\rho U_{bulk}^2)$  for Couette flow and $2\overline{\uptau}_w/(\rho (Q/L_y)^2)$ for pressure-driven flow. The initial flow configurations were chosen from the single-phase flow simulations at $\Rey_b=500$ for Couette and  $\Rey_b=2300$ for channel flow. The particles were then randomly seeded in the simulation domain, at different volumetric concentrations ($\mathit{\Phi}=5$ or $10\%$ in Couette and $\mathit{\Phi}=1$ or $5\%$ in channel flow). The two-phase flow simulations were carried out for several hundreds of time units (typically more than $300$), before the Reynolds number was decreased in order to evaluate the transition threshold. The evolution in time of the wall friction coefficient of the suspension flow is shown in figure \ref{fig:Figure_2} for different cases, after the Reynolds number was abruptly decreased. \\

In Couette flow, the Reynolds number was decreased from $500$ separately to $470$, $455$ or $440$. At $\mathit{\Phi}=10\%$, the flow became laminar for the two simulations at $\Rey_b=455$ and $440$. At $\mathit{\Phi}=5\%$, the flow remained turbulent at $\Rey_b=455$. When the Reynolds number was progressively decreased as following: $455 \rightarrow 415 \rightarrow 390 \rightarrow 365 \rightarrow 355$, the transition took place only around $\Rey_{c} \approx 355$. This critical Reynolds number is calculated using the pure fluid viscosity. The effective Reynolds number based on the suspension viscosity is lower if additional viscous dissipation introduced by the rigid particles is accounted for, due to an increase of the flow viscosity  $\nu_{eff}=\nu \eta(\mathit{\Phi}, \Rey)$, with $\eta(\mathit{\Phi}, \Rey)>1$. There are some possibilities to predict $\eta(\mathit{\Phi}, 0)$ from Eiler's fit \citep{stickel2005fluid} and the correction at finite Reynolds number $\eta(\mathit{\Phi}, \Rey)$ at low concentrations (see for example \citet{subramanian2011influence}). The simulations with FCM give access to the increase of the suspension viscosity induced by the particles through the second order term of the multipole expansion also called Stresslet \citep[for the definition, see][]{wang2017modulation}. Moreover, since the particles are not homogeneously distributed, the increase of the suspension viscosity can be obtained from profiles of $\nu_{eff} $ as shown in figure \ref{fig:Figure_5_1}(c). This leads to a critical Reynolds number of the suspension flow $\Rey_{c,s} \approx 312$ for $\mathit{\Phi}=5\%$ which is very close to the value of single-phase flow $\Rey_{c\_C}$. The main conclusion of this test is that in a Couette flow with moderate particle concentration, the particles act mainly as a source of energy dissipation in the flow, and that they do not change significantly the transition threshold if the suspension viscosity was taken into account in the Reynolds number definition. \\

In pressure-driven flow, the initial flow configuration was selected at  $\Rey_b=2300$. Particles were randomly seeded at concentration $1\%$ or $5\%$. A small concentration of finite-size particles is enough to decrease the transition threshold \citep[see][]{matas2003transition, loisel2013effect, yu2013numerical, lashgari2015transition}, keeping in mind that at low to moderate concentration, the threshold decreases when the concentration is increased, in contrast to Couette flow. Figure \ref{fig:Figure_2}(b) shows the temporal evolution of $C_f$ after particles were seeded and the Reynolds number was decreased from $2300$ to $2000$. For $\mathit{\Phi} =1\%$, the flow is fully laminar at $\Rey_b=2000$. However for $\mathit{\Phi} =5\%$, a stable two-sided turbulent flow is sustained at $\Rey_b=2000$ while \citet{jimenez1991minimal} observed that in the miniunit turbulent flow exists only near one wall in a single-phase flow, even at higher Reynolds number ($\Rey_b=2667$). Decreasing $\Rey_b$ from $2300$ to $1700$ and then to $1500$, the flow becomes laminar at $\Rey_c=1500$ (which corresponds to $\Rey_{c,s}=1315$ based on Eiler's fit whereas $\Rey_{c,s}=1150$ based on \ref{fig:Figure_5_1}(c)). A significant drop of the transition threshold ($\Rey_{c\_P}\sim2200$ for single-phase flow) is observed although the effective viscosity has increased. \\


\subsection*{Influence of particles on the flow stability}

\begin{figure}
  \centerline{\includegraphics[trim = 0mm 0mm 0mm 0mm,width=13.0cm]{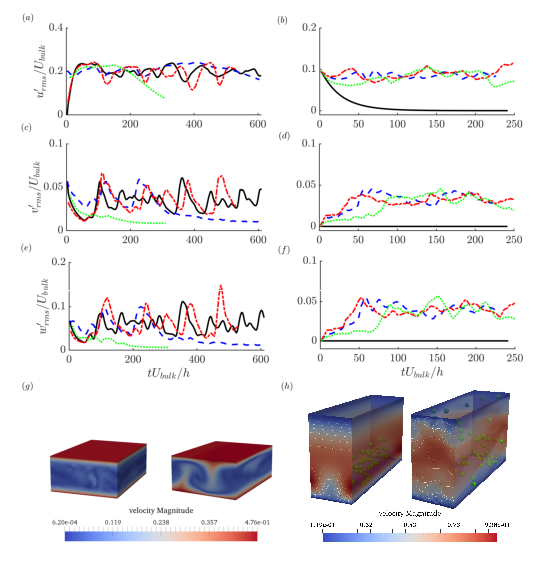}}
  \caption{Particle effect on flow stability. \underline{\textit{Left panel}}: Couette flow starting from a fully turbulent regime. The turbulent state is stable under single-phase condition, even when the streamwise velocity perturbations are suppressed, the flow recovers its fully turbulent nature. Adding particles damp the velocity fluctuations and make the flow laminar. \figline $\Rey_b=430$, single-phase flow removes $u'$; \textcolor{blue}{\figldash} $\Rey_b=430$, $\mathit{\Phi}=5\%$ and $L_y/d=10$; \textcolor{red}{\figddash} $\Rey_b=455$, single-phase flow removes $u'$; \textcolor{green}{\figsdash} $\Rey_b=455$, $\mathit{\Phi}=10\%$ and $L_y/d=20$. \underline{\textit{Right panel}}: channel flow starting with a flow distribution according to (\ref{eq:artificial_1}) where an artifical streak is initially imposed. The single-phase flow tends towards the laminar state at $\Rey_b=2600$ which is above the laminar-turbulent transition. Adding small number of particles in the flow triggers the transition to turbulence. \figline single-phase flow; \textcolor{blue}{\figldash} $\mathit{\Phi}=0.5\%$ and $L_y/d=16$; \textcolor{red}{\figddash} $\mathit{\Phi}=0.75\%$ and $L_y/d=16$; \textcolor{green}{\figsdash} $\mathit{\Phi}=0.5\%$ and $L_y/d=20$. }
\label{fig:Figure_3}
\end{figure}

As will be further discussed in \S  \ref{subsec:Particle_distribution}, particles tend to accumulate in the large scale vortex regions in Couette flow and in low-speed streak regions in channel flow. In order to understand how particles enhance or reduce the flow stability from their preferential spatial distribution, we performed some simulations with specific flow configurations. \\

The first test was done in Couette flow and it was inspired from the study of \citet{hamilton1995regeneration}. When the streamwise velocity perturbations were removed, while the linear streamwise velocity profile and streamwise vortices were maintained from a fully turbulent simulation, the authors observed that the flow evolved again to the fully turbulent regime. In a similar way, we considered a streaks free turbulent flow as initial configuration, a snapshot from a steady single-phase turbulent Couette flow at $\Rey_b=500$, where the amplitude of large-scale x-independent streak achieves one of its peaks (the intensity of the large-scale streaks in miniunit is expresses as $M(0,\beta)$ which is based on modal analysis of the flow fluctuating energy written in (\ref{eq:fft_2d})). \\

Then the Reynolds number was abruptly decreased for single-phase flow to $455$ and $430$ in two separate simulations. The temporal evolution of R.M.S velocity fluctuations shown in figures \ref{fig:Figure_3}(a, c, e) as well as the mean velocity profile (not shown here) suggest that despite the initial destabilization, the flow recovers its fully turbulent features after $200$ time units. Figure \ref{fig:Figure_3}(g) shows the contours of velocity magnitude for single-phase flow, the left one is taken at $t=0$ after removing $u'$ and its evolution after $500$ time units can be seen in the right figure. The flow field plotted after nearly $5$ regeneration cycles cannot be distinguished from the initial fully turbulent flow. Therefore the streamwise vortices, that were initially maintained, were strong enough to generate streaks through the lift-up effect and resumed the regeneration cycle.\\

The main effect of the presence of particles in the Couette flow was stabilizing in nature. When we added small particles ($L_y/d=20$) at $\mathit{\Phi}=10\%$ and decreased the Reynolds number down to $455$, without removing the streamwise velocity perturbations, the flow became laminar. Adding larger particles ($L_y/d=10$) at $\mathit{\Phi}=5\%$ and decreasing the Reynolds number to $430$, the flow velocity fluctuations were significantly damped, and the flow became laminar while staying quasi-turbulent for around $6$ regeneration cycles. These observations suggest that the particles were mainly enhancing dissipation in the flow.  \\


The test on flow stability for the channel flow configuration has been done using an artificial finite-amplitude low-speed streak that was supplemented to a mean flow profile corresponding to wall-bounded turbulence. We used the same base flow as \citet{schoppa2002coherent} who studied the streak transient growth mechanism in a two-dimensional streak configuration. The base flow is:

\begin{equation}\label{eq:artificial_1}
u(y,z)=U_0(y)+(\Delta u/2)cos(\beta_s z)g(y)
\end{equation}
in the streamwise direction, and $v=w=0$ in the wall-normal and spanwise directions. $U_0(y)$ is the mean velocity and $g(y)$ is an amplitude function which satisfies the no-slip condition at $y=0$ and localizes the streak velocity defect at a single wall. A `single-sided' turbulent mean velocity profile is imposed, analogous to that observed in minimal channel turbulence \citep{jimenez1991minimal}, with a parabolic profile $U_{lam}$ in the laminar top half of the channel, and a turbulent Reichardt profile $U_{turb}$, that respects the  near-wall turbulence statistics, in the bottom half:

\begin{equation}\label{eq:artificial_2}
U_0(y)=\left\{
		\begin{array}{ll} 
        U_{lam}=U_c[1-((y/h)-1)^2],~y_m \leq y \leq 2h   \\
		U_{turb}=u_\ast[2.5ln(1+0.4y/\delta)+7.8(1-e^{(-y/11\delta)}-\frac{y}{11\delta} e^{(-y/3\delta)})], \\
		~~~~~~~~~~0\leq y < y_m
  		\end{array} \right.
\end{equation}

The friction velocity $u_\ast=\sqrt{\uptau_w/\rho}$ and viscous length scale $\delta=\nu/u_\ast$ are calculated using a wall shear stress estimated from Dean's empirical correlation ($C_f \equiv 2\tau_w/\rho\overline{u}^2=0.073\Rey_b^{-0.25}$) in a fully turbulent channel flow. For a given flow rate $Q$, this leads the friction velocity to be $u_\ast=Q\sqrt{0.0365(Q/\nu)^{-0.25}}/(2h)$.  The two profiles $U_{turb}$ and $U_{lam}$ are matched at a wall normal distance $y_m$ in the turbulent half, with $y_m$ and $U_c$ determined so that the mean flow velocity and vorticity are continuous at the matching point, i.e. $U_{lam}(y_m)=U_{turb}(y_m)$ and $dU_{lam}/{dy}\mid_{y=y_m}=dU_{turb}/{dy}\mid_{y=y_m}$. Consequently, at $\Rey=Q/\nu=2600$, $y_m=0.918h$ and $U_c=1.2Q/(2h)$. \\

The function $g(y) \sim y \cdot e ^{-\eta y^2}$ is accounting for streak velocity defect, it has been normalized to unity with $\eta$ specified such that the streak velocity defect $\Delta u$ and normal vorticity $\Omega_y \mid_{max} = \beta_s \Delta u /2$ exhibit a plateau in the range $y^+=10-30$, consistent with the observed lifted streaks and $\omega _{y,rms}$ statistics. Note that the amplitude function $g(y)$ in (\ref{eq:artificial_1}) determines the strength of the local streak upper bound $u(y)$ shear layer (e.g. local maxim of $\partial u/ \partial y$) residing on the crest of the lifted streak. Instability growth rates for the dominant sinuous modes are found to be relatively insensitive to the strength of this shear layer and hence to the amplitude function $g(y)$.  The value $\eta=20$ was used similarly to \citet{schoppa2002coherent}. The streak spanwise wavenumber $\beta_s$ in (\ref{eq:artificial_1}) is chosen as $2\pi/\beta^+_s=100$, corresponding to the well-accepted average spanwise spacing of adjacent low-speed streaks observed in many experimental and numerical studies. \\

Figure \ref{fig:Figure_3}(h)-left shows the velocity magnitude contours of the flow according to equation \ref{eq:artificial_1}. \citet{schoppa2002coherent} found that this single-phase flow is stable and the energy of the artifical streaky perturbation will vanish in time due to viscous dissipation. It is necessary for growth a spanwise perturbation following a sinuous profile in the flow direction. In the absence of such spanwise initial coherent motion, figures \ref{fig:Figure_3}(b, d, f) confirm that the perturbation (\ref{eq:artificial_1}) is damped over time when the Reynolds number is equal to $2600$ (above the transition threshold). \\

Unlike the Couette flow test, the particles in this particular channel flow were not seeded throughout the entire domain. Of course this would lead the flow to undergo the transition to turbulence. Instead, we seeded a small number of particles only in the low-speed artificial streak region ($u(y,z)/U_{bulk} \leq 1.5$) keeping the flow Reynolds number $\Rey_b=2600$. Two different local concentrations (particles-to-streak volume) were considered ($\mathit{\Phi}=0.5$ and $0.75\%$ for the case of $L_y/d=16$ and $\mathit{\Phi}=0.5\%$ for the case of $L_y/d=20$). In all cases, the particle presence triggered the transition to turbulence (this can be evidenced by the level of the R.M.S velocity signals), and the particles were found after $\sim 100$ time units spread all over the simulation domain. The figure \ref{fig:Figure_3}(h)-right shows the contours of velocity magnitude for suspension flow in the case of $L_y/d=16$ with $\it{\Phi}=0.75\%$ after $250$ time units. We can observe a quasi fully-turbulent state at $\Rey_b=2600$ (instead of the one-side wall turbulence observed in single-phase flow noted by \citet{jimenez1991minimal}). \\

\begin{figure}
  \centerline{\includegraphics[trim = 0mm 0mm 0mm 0mm,width=13.0cm]{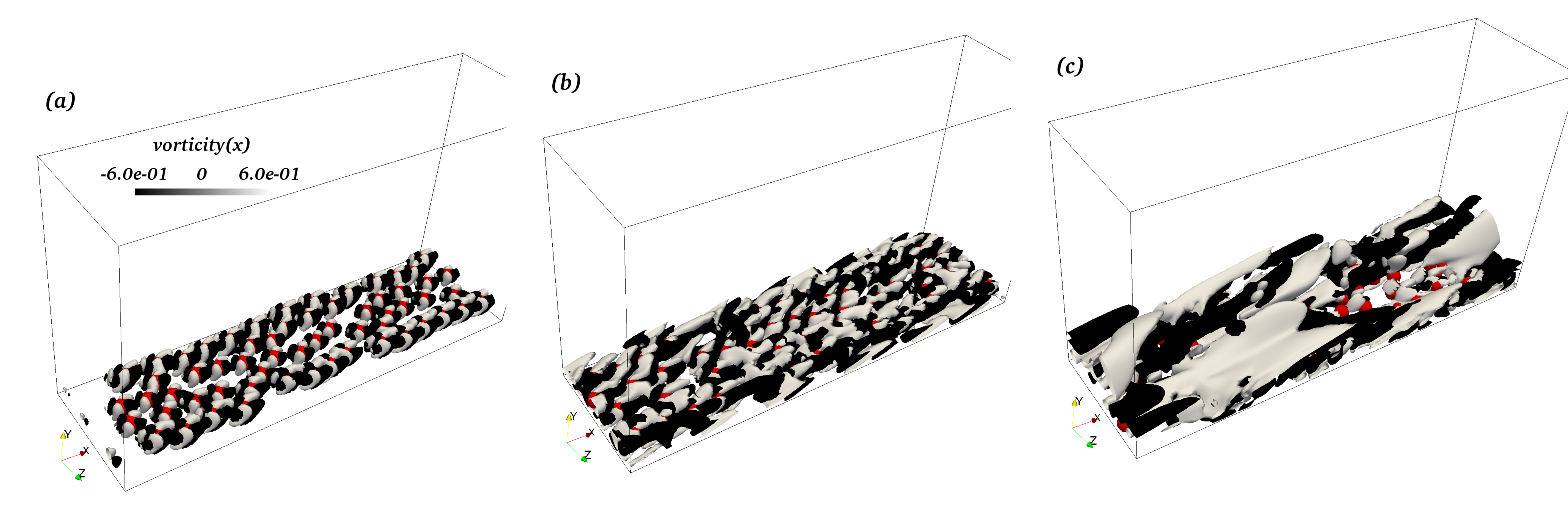}}
  \caption{Flow vorticity induced by a layer of particles seeded in the plane $y/d=0.8$ (near the wall) in the same flow configuration as \ref{fig:Figure_3}(h). The total volume concentration is $\mathit{\Phi}=0.5\%$, and the size ratio $L_y/d=16$ is used. The snapshots are taken at three time instants $t=0.2$, $2.7$ and $29.5$ (scaled by $h/U_{bulk}$) which correspond to $t^+=2$, $27$ and $295$ (scaled by $\nu/u_{\uptau}^2$).}
\label{fig:Figure_4}
\end{figure}

The influence of the particles on the transition is also illustrated in figure \ref{fig:Figure_4} showing the temporal evolution of the streamwise vorticity generated by the particles. The initial condition is equivalent to figure \ref{fig:Figure_3}(h), except that $60$ particles were initially seeded in a plane parallel to the wall (instead of being located in the artificial streak).  At the first instants (figure \ref{fig:Figure_4}(a)), streamwise vorticity is generated around finite-size particles due to the secondary flows occurring at finite $Re_p$. As time goes on and particles move, this streamwise vorticity is stretched in the streamwise direction as in figure \ref{fig:Figure_4} (b). Furthermore, these structures are tilted due to the mean shear through the streamwise vorticity generation term $-({\partial w}/{\partial x})({\partial u}/{\partial y})$ (explained in \S\ref{sec:regeneration_cycle}) which is large near the wall. They further interact with each other to form larger scale streamwise vortical structures as shown in figure \ref{fig:Figure_4} (c). Clearly the generated vortical structures are comparable to the near wall vorticity layers induced by large scale vortices, which are the essential ingredients of the regeneration cycle for channel flow.

\section{Modulation of the turbulent flow energy}\label{sec:energy}
In this section, we show that particles modulate the flow energy in a channel more strongly than in a Couette flow. Since flow modulation is partly related to particle spacial distribution, the latter will be first discussed. 

\subsection{Particle dispersion}\label{subsec:Particle_distribution}

\begin{figure}
  \centerline{\includegraphics[trim = 0mm 0mm 0mm 0mm,width=12.0cm]{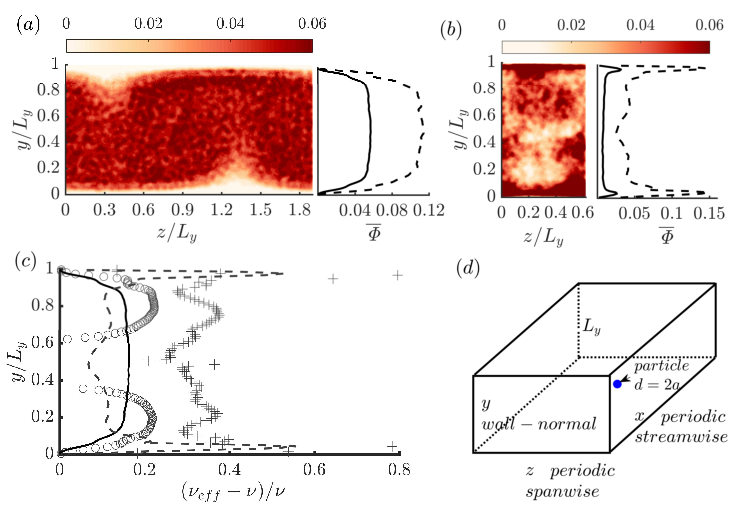}}
  \caption{Particle distribution in $y-z$ plane. $(a)$ and $(b)$ show concentration contours averaged in the streamwise direction, over 80 time units in cases $C500-5$ and $P2600-5$. The corresponding profiles in wall-normal direction are averaged over 500 time units for \figline $C500-5$, \figldash $C500-10$ and \figline $P2600-1$, \figldash $P2600-5$. $(c)$ shows the effective viscosity, based on Eilers fit: \figline $C500-5$; \figldash $P2600-5$, based on local shear stresslet and concentration: circle is $C500-5$; cross is $P2600-5$. $(d)$ is the schematic drawing of flow configurations.}
\label{fig:Figure_5_1}
\end{figure}

Figures \ref{fig:Figure_5_1} (a, b) show the average particle distribution over the cross-section plane. The concentration contours are averaged over $80$ time units ($h/U_{bulk}$) whereas the concentration wall-normal profiles were averaged over $500$ time units. The maximum concentration is located in the core region of the Couette flow, whereas two peaks can be observed near the walls of the channel flow. The average concentration profiles are the result of a balance between the lift force on the finite-size particles, which is oriented towards the center in the Couette flow \citep{ho1974inertial} and towards the walls in the channel flow \citep{asmolov1999inertial}, the hydrodynamic repulsion from the wall and the shear-induced turbulent diffusion. \\

\begin{figure}
  \centerline{\includegraphics[trim = 0mm 0mm 0mm 0mm,width=13.0cm]{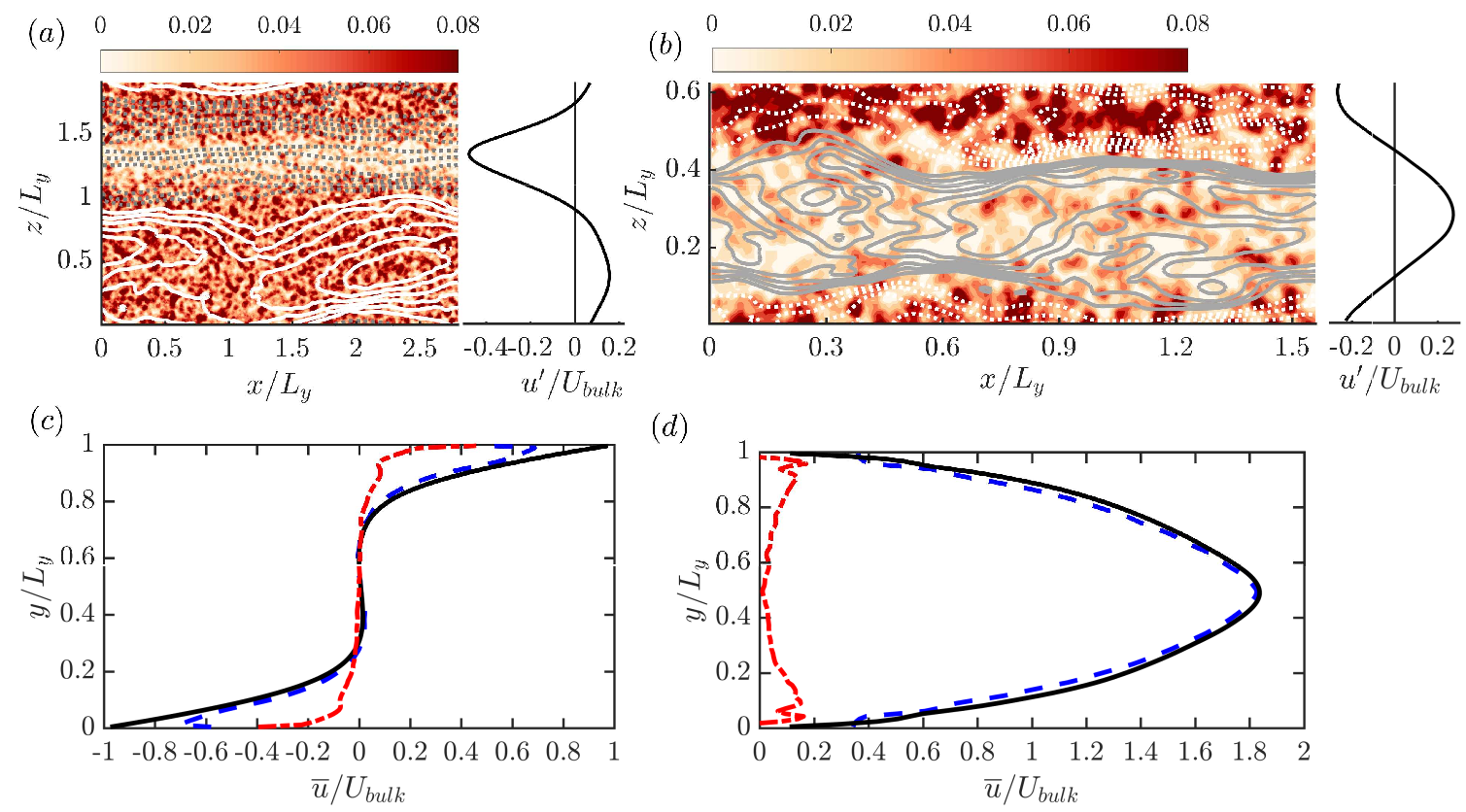}}
  \caption{Particle distribution $x-z$ plane. $(a)$ and $(b)$ are the concentration contours taken at $y/L_y=0.2$ in $x-z$ plane averaged over $80$ time units for $C500-5$ and $P2600-5$, the isolines show $u'/U_{bulk}$ in $x-z$ plane where dashed line stands for negative and solid line shows positive, the interval is $0.04$ in $(d)$ and $0.03$ in $(e)$. The profiles of $u'/U_{bulk}$ averaged in streamwise are also plotted on the right side of $(a)$ and $(b)$, separately. $(c)$ and $(d)$ are the mean velocity of \figline fluid phase, \textcolor{blue}{\figldash} particle phase and \textcolor{red}{\figddash} difference between fluid phase with particle phase.}
\label{fig:Figure_6}
\end{figure}

The instantaneous spatial distribution of particles is shown together with the streamwise velocity fluctuation contours in the $x-z$ plane (figures \ref{fig:Figure_6} (a, b)). These figures show a strong correlation between the particle spatial distribution and the flow coherent structures. In Couette flow, particles are pumped away from walls by turbulent ejection and towards the wall by the sweep events. On average, they are more present in the sweep and core regions in Couette flow. In channel flow, the particles are accumulating in the ejection region, near the wall. \textcolor{black}{As a consequence, the mean velocity of fluid phase should be smaller than the particle phase if the particles are more in the high-speed streaks whereas the mean velocity of fluid phase should be higher than the particle phase if the particles are more in the low-speed streaks. This can be found in figures \ref{fig:Figure_6} (c, d), the shapes of mean velocity profiles of fluid phase and particle phase are similar and the difference between fluid phase with particle phase is clearly negative in Couette flow (near the bottom wall) whereas it is positive in channel flow.}\\

This can be understood as follows: the inertial lift force drives the particles to be preferentially located near the walls, where high and low speed flow regions are encountered. \textcolor{black}{In wall-normal direction, the particles are ejected away from the walls by the ejection event whereas the particles move towards to the walls by the sweep event. Due to the continuity in spanwise direction, sweeps are regions of spanwise divergence near the wall which drives the particles to leave the high speed (sweep) region towards low speed (ejection) region.} \\

\subsection{Quadrant analysis of velocity rms}

\begin{figure}
  \centerline{\includegraphics[trim = 0mm 0mm 0mm 0mm,width=13.0cm]{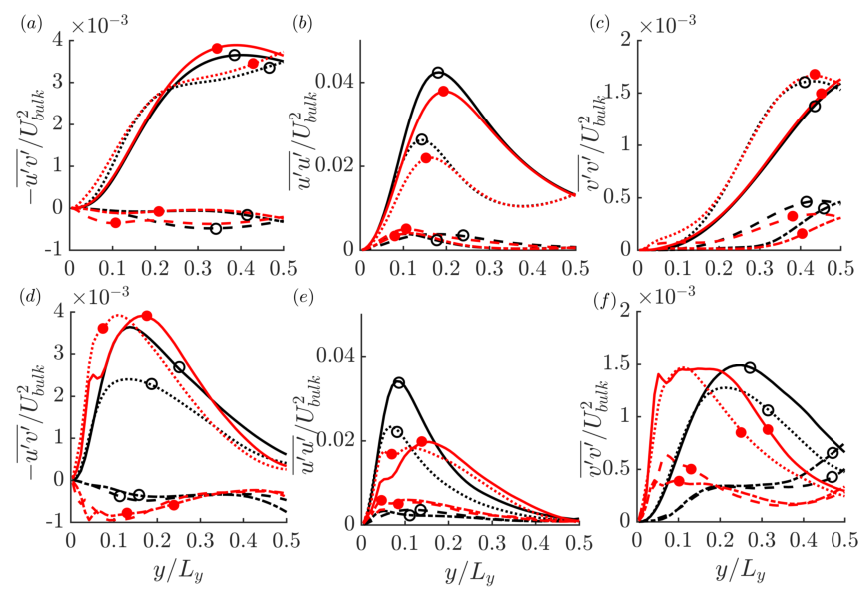}}
  \caption{Velocity rms of single and two phase flows in different quadrants. For ($a\rightarrow f$): the line style denotes the quadrant, \figldash $Q1$, \figline $Q2$, \figddash $Q3$, \figsdash $Q4$. The symbol refers to the flow concentration, circle for single-phase flow and disc for $5\%$ concentration. $(a-c)$: Couette flow. (d-f): channel flow. $(a,d)$, $(b,e)$ and $(c,f)$ show $-\overline{u'v'}$, $\overline{u'u'}$ and $\overline{v'v'}$, respectively.}
\label{fig:Figure_7}
\end{figure}

The coherent structures are the major contributions to the Reynolds shear stress. They play an essential role in the $active$ motion of wall turbulence. According to the quadrant analysis, the flow fluctuations can be divided into $Q1(u'>0,v'>0)$, $Q2(u'<0,v'>0)$, $Q3(u'<0,v'<0)$ and $Q4(u'>0,v'<0)$. $Q2$ and $Q4$ correspond to the ejection and sweep events respectively, $Q1$ and $Q3$ contain the outward and inward interactions respectively (see a recent review on quadrant analysis by \citet{wallace2016quadrant}). We give details on the impact of particles on the different Reynolds stress components, by considering separately the different contributions according to the quadrant analysis. \\

The rms velocity profiles are displayed in figure \ref{fig:Figure_7}, according to the quadrant analysis, both for single-phase and suspension flow at $\mathit{\Phi}=5\%$. The Reynolds stress components are not significantly influenced when particles are present in Couette flow at $5\%$ (figures \ref{fig:Figure_7}(a-c)). However in channel flow, profiles of the Reynolds shear stress (figure \ref{fig:Figure_7}(d)) reveal that the particles enhance significantly the shear stress in the sweep ($Q4$) part of the logarithmic region, where $Q2$ and $Q4$ events are dominant. The streamwise Reynolds stress is decreased by the particles, especially in the ejection regions, whereas the wall-normal Reynolds stress is increased in both sweep and ejection regions. The peak of the profiles are also closer to the wall. \\

\begin{figure}
\centering
\putfig{$(a)$}{\includegraphics[width=11 cm]{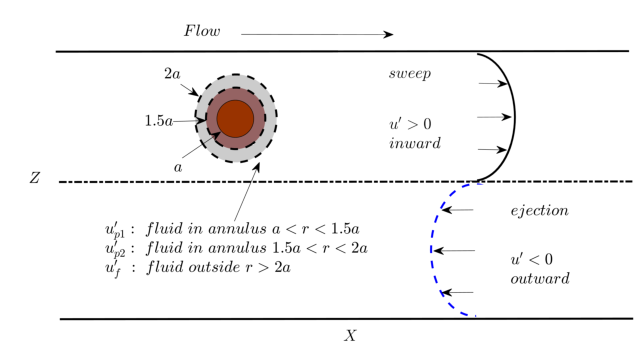}} \quad
\putfig{$(b)$}{\includegraphics[width=6.5 cm]{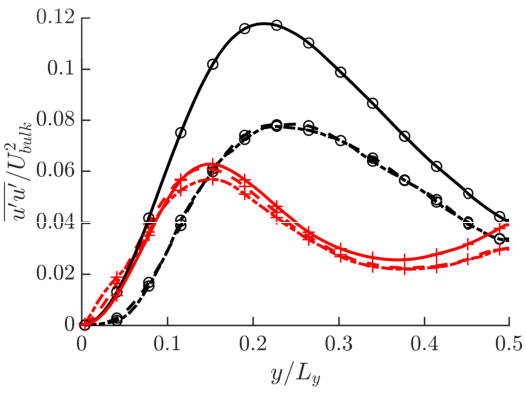}} \quad
\putfig{$(c)$}{\includegraphics[width=6.5 cm]{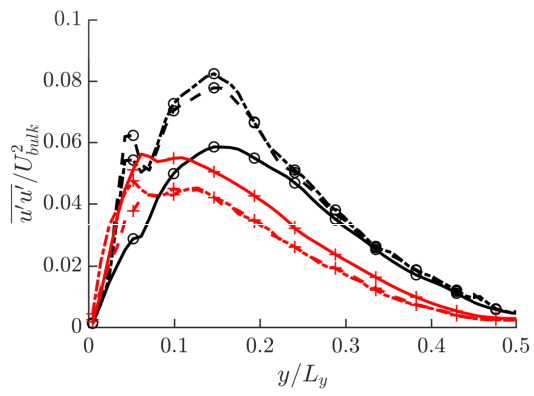}} 
  \caption{ As sketched in figure $(a)$, velocity rms of single and two phase flows in Q2 and Q4 are split into contributions near the particle surface ($\overline{u'_{p1} u'_{p1}}$ withing $a<r<1.5a$ in dot-dashed lines and $\overline{u'_{p2} u'_{p2}}$ withing $a<r<1.5a$ in dashed lines), and far from the particle surface ($\overline{u'_f u'_f}$ in $r>2a$ in lines). Here the circle corresponds to the ejection and the plus to the sweep events. The profiles in $(b)$ are from Couette flow and the profiles in $(c)$ from channel flow.}
\label{fig:Figure_8}
\end{figure}

These turbulent stress modifications suggest a more isotropic turbulence in a channel flow compared to a single-phase flow (in agreement with previous numerical results obtained using the total profiles of the Reynolds stress components \citep[see][]{loisel2013effect,shao2012fully, picano2015turbulent, yu2016finite, fornari2016effect}, the transfer between different directions being promoted by the  Reynolds shear stress. A particular attention is drawn here to the reduction of streamwise Reynolds stress component. \citet{fornari2016effect} related this observation to the fluid squeezed between the layer of particles near the wall and the wall itself. \citet{shao2012fully} associated this with the weakening intensity of the large-scale streamwise vortices which phenomena is not observed in Fourier space as shown later in figure \ref{fig:Figure_10}(d) and figures \ref{fig:Figure_9}(e, f). In order to understand why the Reynolds streamwise stress is decreased, we calculated the velocity fluctuations in fluid and particle regions separately, as sketched in figure \ref{fig:Figure_8}(a). $\overline{u'_f u'_f}$ is averaged within the fluid region located at $r>2a$ relatively to each particle center. $\overline{u'_{p1} u'_{p1}}$ and $\overline{u'_{p2} u'_{p2}}$ is averaged in the neighborhood of the particles $a<r<1.5a$ and $1.5a<r<2a$, respectively. \textcolor{black}{The profiles of $\overline{u'_f u'_f}$, $\overline{u'_{p1} u'_{p1}}$ and $\overline{u'_{p2} u'_{p2}}$ are plotted in different quadrants in figures \ref{fig:Figure_8}(b,c), where we can see there is slight difference between $\overline{u'_{p1} u'_{p1}}$ and $\overline{u'_{p2} u'_{p2}}$. However, from figure \ref{fig:Figure_8}(b) it can be noted that in Q2 the amplitude of the fluid velocity fluctuations around the particles are smaller than the fluid away from the particles in Couette flow whereas it is opposite in channel flow as shown in figure \ref{fig:Figure_8}(c). The perturbation of the fluid around the particles are stronger than the fluid away from the particles in channel flow. This is contradictory with an intuitive explanation that the streamwise velocity rms of the suspension flow reduction (figure \ref{fig:Figure_7}(e)) is due to the lagging of particles in the flow. Even though we do not find the large-scale streamwise vortices are weakened in channel flow, however the amplitude of the large-scale streaks are stabilized as shown in figures \ref{fig:Figure_13}(c, d). Therefore, the reduction of streamwise velocity rms of the suspension flow is due to the stabilized amplitude of the large-scale streaks, but not the weakening intensity of large-scale streamwise vortices.} \\

\subsection{Energy spectra}
\begin{figure}
  \centerline{\includegraphics[trim = 0mm 0mm 0mm 0mm,width=13.0cm]{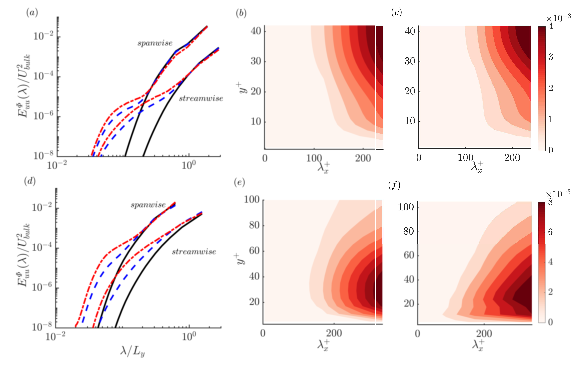}}
  \caption{ Top panels: Couette flow; Bottom panels: channel flow. $(a)$ and $(d)$ show the one-dimensional streamwise and spanwise wavenumber energy spectra of the streamwise velocity $E_{uu}$ averaged in the wall-normal direction. In $(a)$ \figline $C500-0$; \textcolor{blue}{\figldash} $C500-5$; \textcolor{red}{\figddash} $C500-10$. In $(d)$ \figline $P2600-0$; \textcolor{blue}{\figldash} $P2600-5$; \textcolor{red}{\figddash} $P2600-10$. Contour figures show the two-dimensional contours of the energy spectra. $(b)$ and $(e)$: single-phase flow with $C500-0$ and $P2600-0$. $(c)$ and $(f)$: two-phase flow with $C500-5$ and $P2600-5$.}
\label{fig:Figure_9}
\end{figure}

The average streamwise energy spectrum $E^ \Phi _{uu}$ is plotted in figures \ref{fig:Figure_9}(a, d) as a function of both streamwise and spanwise wavelengths, for both suspension and single-phase flows. One advantage of FCM is that particles are represented in the fluid equations by smooth Gaussian envelope forcing, making this method well suited for spatial Fourier analysis of mixture flow. \\

Figures \ref{fig:Figure_9}(a, d) display the energy spectra in both flow configurations and for both streamwise and spanwise wavenumbers averaged over the whole domain. Particles strengthen the energy of the flow structures at intermediate scales especially in the streamwise direction of the channel flow. Moreover, particles hardly affect the energy of the large scale structures of wavelength $3h<\lambda_x<5.7h$ in Couette flow whereas particles increase the energy contained in $h<\lambda_x<3.1h$ for channel flow. Note that the energy of fluctuations at small scales in two-dimension ($\lambda_x<d_p$ and $\lambda_z<d_p$) might be over-estimated by the numerical method. However, the energy contained in these scales obtained from two-dimensional model analysis, is $10^{-7}\%$ (resp. $10^{-4}\%$) of the energy of the largest scales in Couette (resp. channel) flow, and therefore the errors introduced from small scales on the analysis of energy spectra can be neglected. 

The energy spectra in the streamwise direction are plotted as a function of the wall-normal position in figures \ref{fig:Figure_9}(b, c, e, f). It is interesting to note in figures \ref{fig:Figure_9}(b, c) that the most energetic large scale motion in Couette flow is found in the range $20<y^+<40$ which corresponds to the buffer layer region. The extent of the most energetic eddies is slightly shrinked towards the Couette center by the particle presence. \\

In channel flow the most energetic flow structures ($h<\lambda_x<3h$) are located at $10<y^+<50$ as shown in figures \ref{fig:Figure_9}(e, f). Contrary to Couette flow, finite-size particles in channel flow enhance the strength of moderate streamwise vortices in comparison with single-phase flow. The energy of these streamwise vortical structures subsequently increases the flow velocity gradient near the wall as will be shown in figure \ref{fig:Figure_10}. The energy modulation near the walls ($y^+<20$) is due to the interaction of the particles with the streaks rather than their interaction with the large scale vortices. Two indirect evidences may support this conclusion. First, particles in Couette flow do not generate significant modulation near the walls ($y^+<20$) in figure \ref{fig:Figure_9}(c) when particles are in large scale vortices. Second, in channel flow we also observe the generation of vortical energetic structures when particles are seeded in the bottom wall only with artificial streaks (as shown in figure \ref{fig:Figure_4}(c)). \\

\section{Modification of the regeneration cycle by particles} \label{sec:regeneration_cycle}

The period of the regeneration cycle can be identified from the low frequency evolution of the friction coefficient signal or Reynolds shear stress. First the effect of the particles on the intermittency of the flow will be characterized by considering the fluctuation in time of the friction coefficient and Reynolds shear stress. Second their impact on the successive sub-processes of the regeneration cycle will be qualitatively detailed in the following sub-sections, considering $(I)$ the lift-up mechanism yielding streak formation, $(II)$ the modal analysis of flow velocity fluctuations for its indication on the correlation between the x-dependent ($m>0$ and $n \geq 0$) and x-independent ($m=0$ and $n>0$) streaks and $(III)$ the vorticity stretching and vortex regeneration.

\subsection{Wall friction coefficient and streamwise vorticity}

\begin{figure}
  \centerline{\includegraphics[trim = 0mm 0mm 0mm 0mm,width=13.0cm]{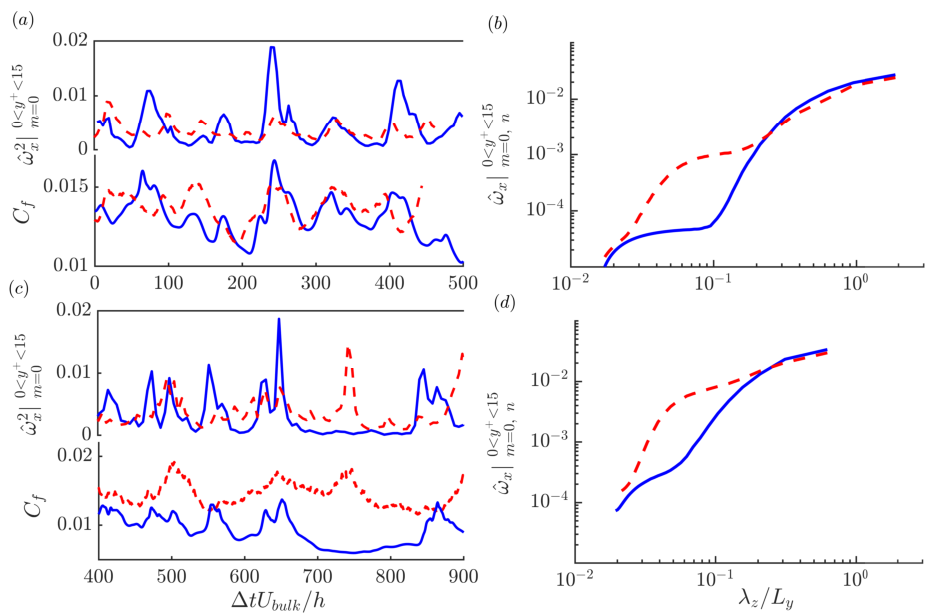}}
  \caption{Simultaneous temporal evolution of the friction coefficient $C_f$ and the near wall streamwise vorticity, in Couette flow ($(a)$ and $(b)$), and channel flow ($(c)$ and $(d)$). $(a)$ and $(c)$ plot the summation of the amplitude square of x-independent vortices ($m=0$) in different spanwise wavenumbers ($1\leq n \leq N_z/2$) and integrated in the near wall region ($y^+<15$). $(b)$ and $(d)$ show x-independent vortices ($m=0$) in the near wall region ($y^+<15$) as a function of spanwise wavelength ($2L_z/N_z \leq \lambda_z \leq L_z$). The line style indicates single-phase (solid) and two-phase (dashed line) flows. In $(a)$ and $(b)$: \textcolor{blue}{\figline} $C500-0$; \textcolor{red}{\figldash} $C500-5$, in $(c)$ and $(d)$: \textcolor{blue}{\figline} $P2600-0$; \textcolor{red}{\figldash} $P2600-5$.}
\label{fig:Figure_10}
\end{figure}

The friction coefficient is a dimensionless measure of the wall shear stress. The temporal evolution of the friction coefficient is displayed in figure \ref{fig:Figure_10}. For Couette flow, the average friction coefficient and temporal fluctuations are slightly increased by the presence of particles, whereas for channel flow the increase of the average friction coefficient is more significant, and the fluctuation amplitude is slightly reduced. The increase of friction coefficient cannot be exclusively related to the increase of the suspension effective viscosity, since the ratio of the time averaged friction coefficient of the suspension to single-phase flow is around $1.4$, whereas the viscosity enhancement is only $1.14$ based on Eilers fit. Recent work from \citet{costa2016universal} provided a theoretical prediction of the total suspension drag. Predicted $\Rey_\uptau$ is $103$ based on suspension viscosity from figure \ref{fig:Figure_5_1}(c) where $\Rey_\uptau$ is $109$ based on DNS in table \ref{tab:Table_1}. \\

\citet{jimenez1991minimal} have explicitly shown that the maximum (in time) of the wall shear stress is synchronous with the maximum near-wall vorticity ($0<y^+<10$). Using 2D numerical simulations (neglecting the variation in the streamwise direction), \citet{orlandi1994generation} showed that the transport of fluid from longitudinal vortices to the high and low speed streaks is the origin of the higher wall friction in turbulent layers, especially in the sweep region where high-speed fluid is transported towards the wall. Therefore large scale streamwise vortical structures control the near-wall velocity gradient.
Figures \ref{fig:Figure_10}(a, c) show simultaneously the evolution of wall friction coefficient and the summation over all the spanwise wavenumbers for x-independent vorticity  $\hat{\omega}_x^2|^{~0<y^+<15}_{~m=0} \equiv \sum_{n=1}^{N_z/2}\int_0^{y^+=15}\hat{\omega}_x^2(0,n\beta)dy$, \textcolor{black}{where the mode $(0,n\beta)$ with $n\neq0$ is an x-independent structure. Further the cross correlation between $C_f$ and $\hat{\omega}_x^2|^{~0<y^+<15}_{~m=0}$ at the same time step as is calculated based on (\ref{eq:cross_correlation}), where $R=0.51,~0.49,~0.63~,0.68$ corresponding to $C500-0,~C500-5,~P2600-0,~P2600-5$, respectively. The correlation in time between the variation of the average friction coefficient and the near-wall streamwise vorticity is obvious in all cases.}

\begin{equation}\label{eq:cross_correlation}
R(C_f, \hat{\omega}_x^2|^{~0<y^+<15}_{~m=0}) = \frac{\overline{C_f~(t) \cdot \hat{\omega}_x^2|^{~0<y^+<15}_{~m=0}~(t) }}{rms(C_f)\cdot rms(\hat{\omega}_x^2|^{~0<y^+<15}_{~m=0})}
\end{equation}

The spectra of near-wall streamwise vorticity $\hat{\omega}_x|^{~0<y^+<15}_{~m=0,~n} \equiv \int_0^{y^+=15}\hat{\omega}_x(0,n\beta)dy$ are shown in figures \ref{fig:Figure_10}(b, d) for both single-phase and suspension flows. The ratio of the vorticity at small scales ($\lambda_z/L_y<0.2$) to the largest scale streamwise vortices is much smaller in Couette flow than in channel flow (the ratios are $O(0.01)$ and $O(0.1)$ respectively). For two-phase flows, turbulence becomes more isotropic, mainly because particles inject energy in small scales, which is transferred back to intermediate scales. This is comparable to the work of \citet{elghobashi1993two}, even though the origin of momentum is not the same: in their work it is due to the slip between the phases, whereas in our work it is mainly due to the particle finite-size. The streamwise vorticity is enhanced at low spanwise wavelengths ($\lambda_z/L_y<0.2$ which corresponds to $\lambda_z/d_p<4$) in both configurations. The enhancement is of one order of magnitude in Couette flow, and of two orders of magnitude in channel flow.  \\

\subsection{Reynolds shear stress}

\begin{figure}
  \centerline{\includegraphics[trim = 0mm 0mm 0mm 0mm,width=13.0cm]{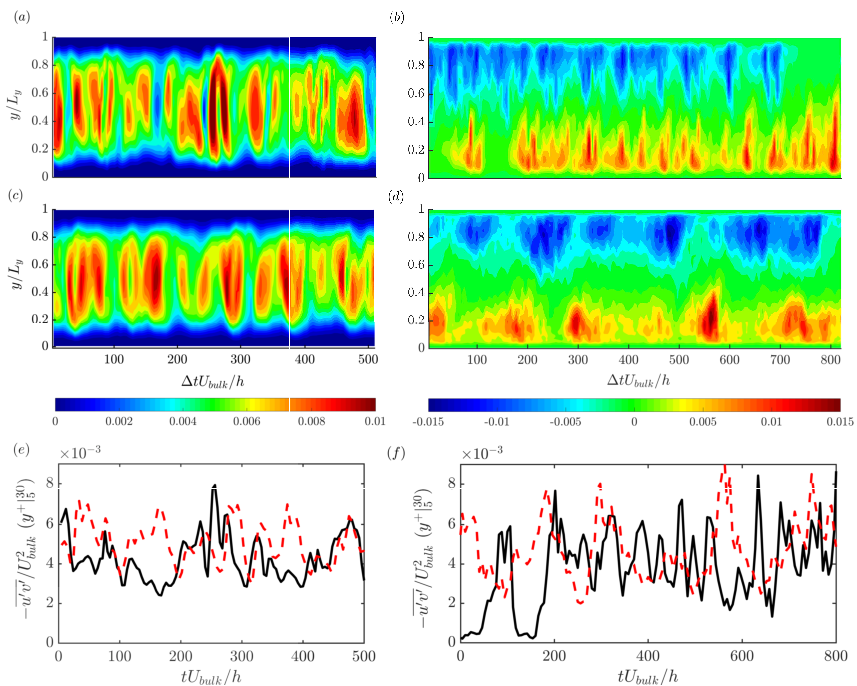}}
  \caption{Spatial-temporal evolution of the Reynolds shear stress ($-\overline{u'v'}/U^2_{bulk}$) averaged in the homogeneous directions (streamwise and spanwise) in $(a-d)$, \textcolor{black}{its average within $5<y^+<30$ is shown in $(e,f)$. Left panel is for Couette flow and right panel is for channel flow. $(a)$ and $(b)$ correspond to single-phase flows whereas $(c)$ and $(d)$ correspond to two-phase flows. In $(e)$, \figline $C500-0$ and \textcolor{red}{\figldash} $C500-5$; In $(f)$, \figline $P2600-0$ and \textcolor{red}{\figldash} $P2600-5$.} }
\label{fig:Figure_11}
\end{figure}

From the investigation of total energy input and dissipation rate, \citet{kawahara2001periodic} evidenced the temporal evolution of spatial structures, in a cyclic sequence consistent with the regeneration cycle proposed by \citet{hamilton1995regeneration}. A strong ejection event is followed by a gradual decrease of intensity over a certain period of time. The maximum (in time) of the Reynolds stress occurs when the dissipation rate is large along the periodic orbit. The quasi-periodicity of the turbulent events can be represented by the spatial-temporal evolution of the Reynolds stress $-\overline{u^\prime v^\prime}(y,t)$ across the Couette gap or channel height, as shown in figure \ref{fig:Figure_11}. Note that the period of turbulent events is of $O(100)$ time units in both flows. It characterizes the time needed for the velocity fluctuations to become uncorrelated in time. The particle Stokes number which can be based on this time scale is very small compared to the one related to the shear. \\

For Couette flow, the maximum of the Reynolds stress is located in the center of the gap. The two walls share one buffer layer and a couple of central large scale vortices, with a strong coupling between the streaks near both walls. The low-speed streak near one wall ejects fluid to the other wall acting there as a high-speed streak. It is revealed by figures \ref{fig:Figure_11}(a, c, e) that neutrally buoyant particles have a negligible effect on both the intensity and intermittency of the Reynolds stress in Couette flow configuration \citep{wang2017modulation}. The channel flow contains log-law region and the central region is ruled by the velocity-defect law.  Figure \ref{fig:Figure_11}(b) shows that the strongest shear stress bursts are located close to the channel walls, and that the frequency of these bursts is of the same order of magnitude as in Couette flow. In the presence of neutrally buoyant particles, the intensity of the Reynolds shear stress is enhanced as shown in figure \ref{fig:Figure_11}(d), and the frequency of these events is decreased as shown in figure \ref{fig:Figure_11}(f). The increase of Reynolds shear stress is closely correlated with the sweep events as indicated in figure \ref{fig:Figure_7}(d), making the friction coefficient and Reynolds shear stress fluctuations synchronous. \\

\subsection{Streak formation: the lift-up mechanism}

\begin{figure}
  \centerline{\includegraphics[trim = 0mm 0mm 0mm 0mm,width=13.0cm]{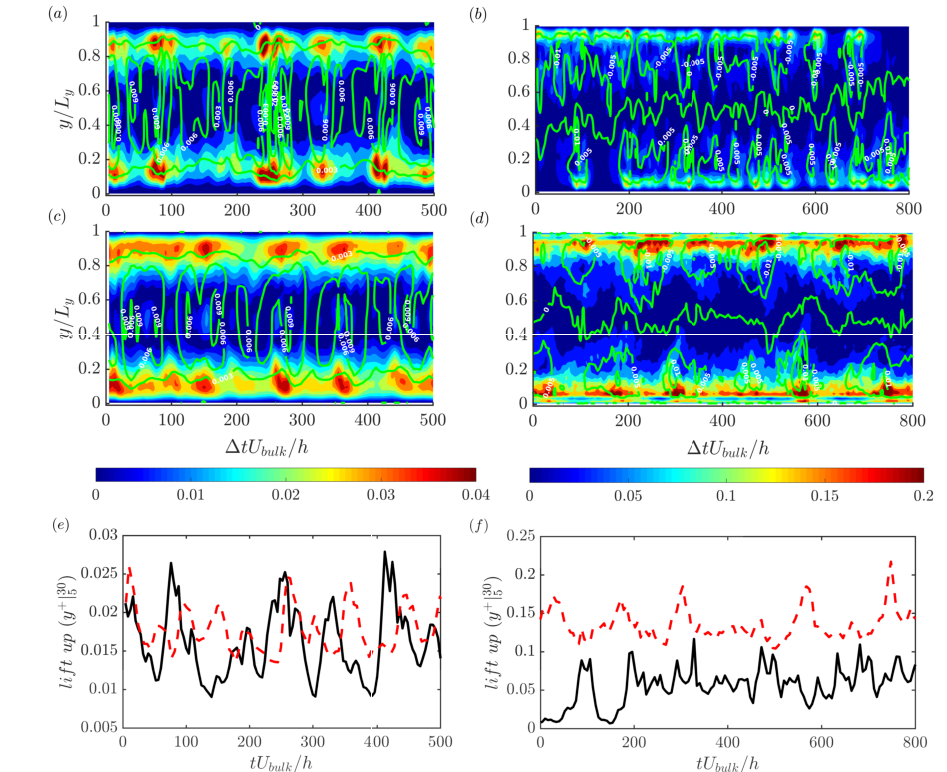}}
  \caption{Spatial-temporal evolution of the lift-up term ($-vd\overline{u}/dy$) scaled by $U^2_{bulk}/h$. The green isoline of the Reynolds shear stress ($-\overline{u'v'}/U^2_{bulk}$) are added on the top of these figures, the interval in $(a)$ and $(c)$ is $0.003$, and the interval in $(b)$ and $(d)$ is $0.005$. Left panel is for Couette flow and right panel for channel flow. $(a)$ and $(b)$ stand for single-phase flows. $(c)$ and $(d)$ stand for two-phase flows. \textcolor{black}{Its average within $5<y^+<30$ is shown in $(e,f)$. $(e)$, \figline $C500-0$ and \textcolor{red}{\figldash} $C500-5$; $(f)$, \figline $P2600-0$ and \textcolor{red}{\figldash} $P2600-5$.}}
\label{fig:Figure_12}
\end{figure}

The streaks form on both sides of a vortex. Low-speed fluid is lifted-up away from the wall by the vortex into a region of higher-speed fluid, producing a low-speed streak, while on the other side of the vortex, high-speed fluid is pushed towards the wall, creating a high-speed streak. \citet{ellingsen1975stability} have shown, using a linear stability analysis that the x-independent streamwise perturbations grow linearly in time as $-v(d\overline{u}/dy)t$ (the so-called lift-up effect), making any shear flow $\overline{u}(y)$ unstable to x-independent (transverse) perturbations. Consequently in shear flows, the main linear mechanism for transient disturbance growth is the lift-up effect that produces high and low speed streaks in the streamwise velocity. 
\citet{bech1995investigation} stated that the inner shear layer is formed through the lift-up of low-speed streaks from the viscous sublayer, then the shear layers are coupled to an instantaneous velocity profile with inflectional character and they have been observed to become unstable and break up into chaotic motion, so called `bursting'.
The lift-up effect or advection was identified as a robust mechanism for generation of the streaky motions both in transitional and turbulent flows \citep{ellingsen1975stability, hamilton1995regeneration, del2006linear}.\\

\citet{klinkenberg2013numerical} have shown that small pointwise inertial particles do affect the transition to turbulence not by altering the lift-up effect but rather by modifying the dynamics of the oblique waves necessary for the streaks regeneration and breakdown. In order to show whether finite-size particles modify the lift-up term, the contours of $-vd\overline{u}/dy$ are displayed in figure \ref{fig:Figure_12} together with the isolines of the Reynolds shear stress (from figure \ref{fig:Figure_11}). The lift-up term is important near the walls in both flow configurations. \textcolor{black}{For Couette flow, the contours shown in figures \ref{fig:Figure_11}(a, c) are not significantly changed by the presence of the particles. From figures \ref{fig:Figure_11}(a, c) which plot the averaged value within $5<y^+<30$, it shows only the strongest effects are nearly same with or without particles whereas the weakest effects are enhanced by the particles. However in channel flow within the buffer layer ($5<y^+<30$), the particles not only significantly enhance the lift-up, but also to let it act continuously (lift up effect is less frequently in suspension flow as shown in figure \ref{fig:Figure_11}(f)).} \\

\subsection{Streak breakdown: Modal decomposition of the fluctuating energy}

\begin{figure}
  \centerline{\includegraphics[trim = 0mm 0mm 0mm 0mm,width=13.0cm]{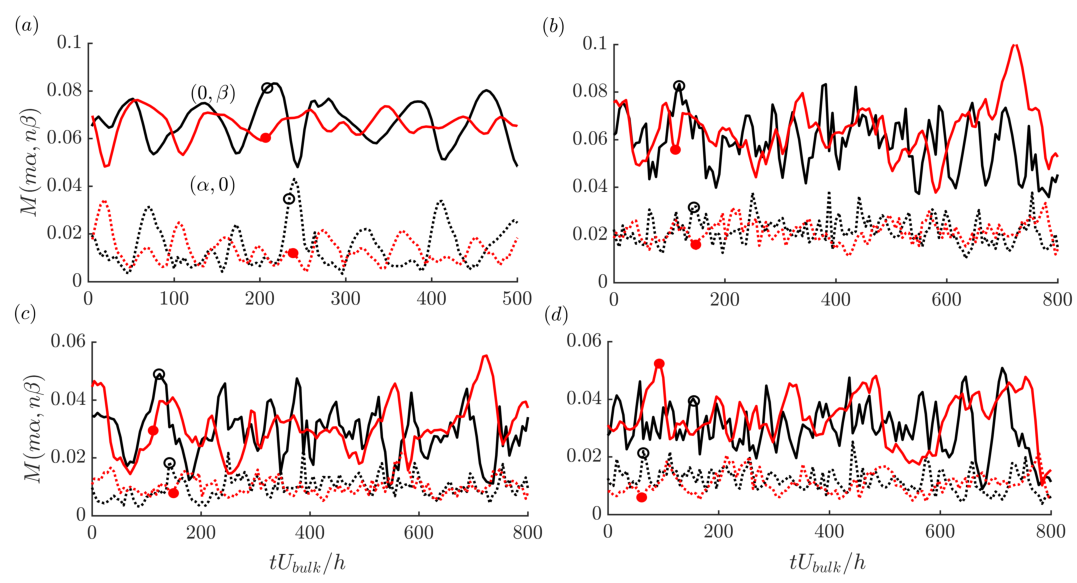}}
  \caption{Modal decomposition as in (\ref{eq:fft_2d}): \figline is mode $M(0,\beta)$; \figsdash is mode $M(\alpha,0)$. $(a)$ Couette flow in the whole domain, $(b)$ Channel flow in the whole domain, $(c)$ and $(d)$ stand for channel flow in the upper half domain and bottom half domain. In $(a)$, black is $C500-0$ and red is $C500-5$. In $(b)$, $(c)$ and $(d)$, black circle is $P2600-0$ and red disc is $P2600-5$.}
\label{fig:Figure_13}
\end{figure}

The subsequent process is the instability of x-independent streaks, the so-called streak breakdown. \citet{hamilton1995regeneration} have shown that it is the instability of the streaks (through a non linear process) which causes breakdown.
We investigated the temporal evolution of the energy contained in the dominant flow fluctuation modes, since it can give evidences on the dynamics of the streak breakdown process, and on the particle modulation of this process. \\



The temporal evolution of the most energetic modes is shown in figure \ref{fig:Figure_13}. In figures \ref{fig:Figure_13}(a, b), (\ref{eq:fft_2d}) is integrated between the two walls ($Y_1=0$ and $Y_2=L_y$), whereas in figures \ref{fig:Figure_13}(c, d), the integration is performed in the vicinity of one single wall which is regarded as an individual shear layer ($Y_1=0 \rightarrow Y_2=L_y/2$ near the bottom wall, and $Y_1=L_y/2\rightarrow Y_2=L_y$ near the upper wall). The quasi-periodic fluctuations of these modes, with period $\sim100h/U_{bulk}$ for Couette flow, are related to the regeneration cycle. The strongest mode is $M(0,\beta)$ which corresponds to x-independent streaks. As a general trend, neutrally buoyant particles decrease the amplitude of the fluctuations of this mode, whereas they do not have significant impact on its period, which is related to the regeneration cycle. However in channel flow, it can be noted that both $(0,\beta)$ mode and $(\alpha,0)$ mode are of the same strength and period compared between single with two-phase flow. \\

In Couette flow, one can note the relation of the intermittency of modes $(0,\beta)$ (x-independent) and $(\alpha,0)$ (x-dependent), when integrated over the entire gap. \textcolor{black}{The cross correlation at the same time step is calculated based on (\ref{eq:cross_correlation}), which is $-0.64$ for $C500-0$ and $-0.59$ for $C500-5$}. The peaks of $M(0,\beta)$ correspond to instants at which the streaks have the least x-dependence. As the streaks become wavy, $M(0,\beta)$ decreases, while the energy of $M(\alpha,0)$ (the fundamental mode in $x$ direction with no spanwise variation) sharply increases. The other $(\alpha,n\beta),n \neq 0$ modes can hardly be distinguished. Breakdown occurs while $M(0,\beta)$ reaches a minimum. The amplitude of both mode fluctuations is slightly damped by the particle presence as shown in figure \ref{fig:Figure_13}(a).\\

For channel flow, figure \ref{fig:Figure_13}(b) shows higher frequency fluctuations than in Couette flow, and less correlation between $(\alpha,0)$ and $(0,\beta)$ modes, when integrated over the whole domain. This is due to two coexisting shear layers (one at each wall) which are relatively independent of each other (turbulent mixing is weak in the core region between the two shear layers at low Reynolds number). When the modal energy is integrated over half of the channel section shown in figures \ref{fig:Figure_13}(c, d), one can notice a stronger correlation between $(0,\beta)$ mode and $(\alpha,0)$ mode, like in Couette flow, although it is less pronounced in channel flow. The particles do not have a strong effect on the temporal evolution of these modes, suggesting that particles do not significantly alter the breakdown process.\\

\subsection{Vortex regeneration: Non linear interaction and vortex stretching}

\begin{figure}
  \centerline{\includegraphics[trim = 0mm 0mm 0mm 0mm,width=13.0cm]{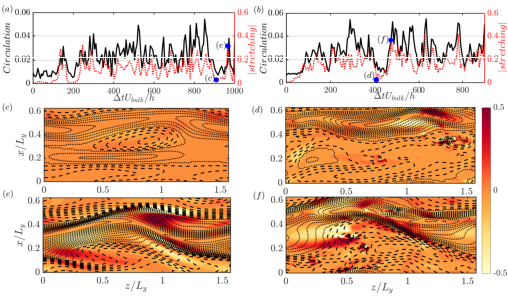}}
  \caption{$(a)$ and $(b)$ show temporal evolution of \figline circulation (integrated over $0.1<y/L_y<0.4$) and \textcolor{red}{\figsdash} spacial average of absolute value of vorticity stretching ($\mid \omega_x\partial u/\partial x \mid$ within $0.1<y/L_y<0.4$ where large scale vortices take place as seen in figure \ref{fig:Figure_5_1}(b)). $(c-f)$ show contours of the streamwise velocity fluctuations $u'/U_{bulk}$ in the snapshot plane at $y/L_y=0.2$, showing the streaks. The interval of isolines is 0.04. \figsdash stands for $u'/U_{bulk}<0$ and \figldash stands for $u'/U_{bulk}>0$. The color contours indicate the stretching term $\omega_x\partial u/\partial x$. Left panels are for single-phase flow $P2600-0$, two points $(c)$ and $(e)$ in $(a)$ are x-independent flow (at the trough of the vorticity stretching instant $\Delta tU_{bulk}/h=900$) and for x-dependent flow (at the peak of the vorticity stretching instant $\Delta tU_{bulk}/h=970$). Right panels are for suspension flow $P2600-1$, two points $(d)$ and $(f)$ in $(b)$ are corresponding to $\Delta tU_{bulk}/h=405$ and $\Delta tU_{bulk}/h=480$. Line styles in $(d)$ and $(f)$ are same as in $(c)$ and $(e)$.}
\label{fig:Figure_14}
\end{figure}

During streak breakdown, a complex set of interactions re-enforces the streamwise vortices, leading to the formation of a new set of streaks, and completing the regeneration cycle. \citet{hamilton1995regeneration} proposed that strengthening the vortices is due to interactions among the $\alpha$-modes, that grow during the streak breakdown. \citet{schoppa2002coherent} suggested that the vortex formation is inherently three-dimensional, with direct stretching (inherent to streak $(x,z)$-waviness) of near-wall $\omega_x$ sheets leading to streamwise vortex collapse. They provided insights into the dynamics of near-wall vortex formation through the equation of inviscid evolution for streamwise vorticity: 

\begin{equation}
\frac{\partial{\omega_x}}{\partial{t}}=-u\frac{\partial{\omega_x}}{\partial x}\underbrace{-v\frac{\partial{\omega_x}}{\partial y}-w\frac{\partial{\omega_x}}{\partial z}}_\textit{advection} +\underbrace{{\omega_x}\frac{\partial u}{\partial x}}_\textit{stretching}+\underbrace{\frac{\partial v}{\partial x}\frac{\partial u}{\partial z}-\frac{\partial w}{\partial x}\frac{\partial u}{\partial y}}_\textit{tilting}
\label{eq:vorticity_dynamic} 
\end{equation}

In fully developed turbulence, the greatest contribution, in magnitude, to the temporal evolution of the vorticity $\partial \omega_x / \partial t$ is related to the tilting term. This is confirmed by our simulations (not shown here). However \citet{schoppa2002coherent} have stated that this term contributes to the thin tail of the near-wall $\omega_x$ layer, and is not responsible for x-independent streamwise vortex formation ($(0,\beta)$ mode in miniunit). Instead, vortex formation is dominated by stretching of streamwise vorticity. The local $\omega_x$ stretching is sustained in time and is mainly responsible for the vortex collapse, whose location coincides with the $+\omega_x \partial u / \partial x$ peak. The meandering of streaks provide the generation of $\partial u / \partial x$. Then direct stretching of positive and negative $\omega_x$ occurs in regions where $\partial u / \partial x$ is generated across the wavy streak flanks during the streak breakdown process. 
The stretching term is active only during the peaks of the cycle when local three-dimensionality is induced after streak breakdown \citep[see][]{jimenez1991minimal}. \\

Figure \ref{fig:Figure_14} shows instantaneous snapshots containing contours of both the streamwise velocity fluctuations (that illustrate the streak shape), as well as the vorticity stretching term. \textcolor{black}{From figures \ref{fig:Figure_14}(a, b), the cross correlation at the same time step is calculated based on (\ref{eq:cross_correlation}), which is $0.576$ for $P2600-0$ and $0.624$ for $P2600-5$. This shows that high flow circulation is synchronized with the appearance of x-dependent flow structures.} The figure is a clear evidence that non-linear processes like streak breakdown, and thereby vortex regeneration, take place in the suspension flow like in the single-phase flow. In our previous paper \citep{wang2017modulation}, we have shown that the streak waviness and vorticity stretching are almost unchanged for Couette flow in the presence of neutrally buoyant particles. However, in channel flow, from figures \ref{fig:Figure_14}(a, b), the vorticity stretching (averaged value is $0.17$ (resp $0.11$) for $P2600-1$ (resp $P2600-0$)), and the circulation (averaged value is $0.025$ (resp $0.02$) for $P2600-1$ (resp $P2600-0$)), are enhanced near the channel walls due to the presence of particles. The snapshot shown in figures \ref{fig:Figure_14}(e, f) corresponding to high vorticity stretching (high flow circulation) shows that the wavy streaks are smaller and more numerous in suspension flow, when compared to the single-phase flow. 
\section{Concluding remarks}\label{sec:Conclusion}

We have studied turbulent suspension flows in plane Couette and pressure-driven (channel) configurations, slightly above the laminar-turbulent transition. Dilute to moderately concentrated suspensions of neutrally buoyant finite-size spherical particles were considered, the particles diameter being twenty times smaller than the Couette gap or channel height. The simulation domain was chosen to ensure a minimal set of coherent flow structures sufficient to sustain turbulence, in both flow configurations respectively. The effect of particles on transition was first examined, both in fully-developed turbulent flow and in artificially perturbed configurations. Particles were found to trigger instability in channel flow whereas they were mainly dissipating energy in the Couette flow configuration due to their finite size. \\

In the turbulent regime, detailed temporal and spatial analysis, in physical and Fourier spaces, were proposed. Particles did not modify significantly the features of plane Couette flow, whereas they had a clear impact on channel flow. The particle spatial distribution was found to be non-uniform over the cross-section. Particles are more present in the core of the large scale rolls ($inactive$ motion) in Couette flow, and in the ejection ($active$ motion) regions in channel flow. This finding is essentially related to wall-normal inertial lift forces (on finite-size particles) that act in opposite directions depending on the flow configuration. \\

Contrary to Couette flow, the accumulation of particles in the $active$ region of turbulence regeneration for the channel flow configuration yielded clear modifications of the flow statistics and dynamical response. Particles accumulated at the wall due to Segr{\'e}-Silberberg effect are ejected by Q2 events. They are populating ejection regions and the low speed streaks. We observed a reduction of streamwise velocity rms and an increase of the wall-normal component. The wall shear stress was also significantly increased because particles had reinforced the activity of larger scale x-independent streamwise vortices near the walls.  \\

The regeneration cycle of wall turbulence has been studied in presence of particles. Despite the universality of wall turbulence, the Couette flow is constituted of a single shear layer whereas channel flow has two shear layers with opposite signs, leading to different flow response to perturbations. The three successive sub-steps of the regeneration cycle were modified by finite-size particles, since they actively contribute to the dynamics of the buffer layer. We observed an enhancement of the lift-up mechanism together with reinforced Reynolds shear stress (although the frequency of burst events was decreased). Vorticity stretching was increased leading to smaller and more numerous wavy streaks for pressure-driven two-phase flow. Thanks to their preferential presence near the walls, particles triggered small scale vortices that were stretched by the shear flow and survived even at Reynolds numbers below the transition limit of single-phase flow. By studying two distinct turbulent flow configurations laden with neutrally buoyant finite-size particles, we were able to show the specific response of turbulent structures and the modulation of the fundamental mechanisms composing the regeneration cycle of near-wall turbulence. \\

\section{Acknowledgement}

This work was granted access to the HPC resources of CALMIP under the allocation 2015 and 2016-P1002 and of GENCI under the allocation x20162b6942. 
We are grateful to COST Action MP1305 on Flowing Matter. Great help from A. Pedrono for technical support on JADIM is also acknowledged.

\bibliographystyle{jfm}
\bibliography{bib/BOOKS_JAB,bib/SINGLE-PHASE,bib/TWO-PHASE_JAB,bib/THESIS_JAB}

\begin{thebibliography}{36}
\expandafter\ifx\csname natexlab\endcsname\relax\def\natexlab#1{#1}\fi
\def\au#1{#1} \def\ed#1{#1} \def\yr#1{#1}\def\at#1{#1}\def\jt#1{\textit{#1}}
  \def\bt#1{#1}\def\bvol#1{\textbf{#1}} \def\vol#1{#1} \def\pg#1{#1}
  \def\publ#1{#1}\def\arxiv#1{#1}\def\org#1{#1}\def\st#1{\textit{#1}}

\bibitem[Asmolov(1999)]{asmolov1999inertial}
{\sc \au{Asmolov, E.~S.}} \yr{1999}  \at{The inertial lift on a spherical
  particle in a plane poiseuille flow at large channel reynolds number}.
  \jt{J. Fluid Mech.}  \bvol{381},  \pg{63--87}.

\bibitem[Bech {\em et~al.\/}(1995)Bech, Tillmark, Alfredsson \&
  Andersson]{bech1995investigation}
{\sc \au{Bech, K.~H.}, \au{Tillmark, N.}, \au{Alfredsson, P.~H.} \&
  \au{Andersson, H.~I.}} \yr{1995}  \at{An investigation of turbulent plane
  couette flow at low reynolds numbers}.  \jt{J. Fluid Mech.}  \bvol{286},
  \pg{291--326}.

\bibitem[Bellani {\em et~al.\/}(2012)Bellani, Byron, Collignon, Meyer \&
  Variano]{bellani2012shape}
{\sc \au{Bellani, G.}, \au{Byron, M.~L.}, \au{Collignon, A.~G.}, \au{Meyer,
  C.~R.} \& \au{Variano, E.~A.}} \yr{2012}  \at{Shape effects on turbulent
  modulation by large nearly neutrally buoyant particles}.  \jt{J. Fluid Mech.}
   \bvol{712},  \pg{41--60}.

\bibitem[Bradshaw(1967)]{bradshaw1967inactive}
{\sc \au{Bradshaw, P.}} \yr{1967}  \at{‘inactive’motion and pressure
  fluctuations in turbulent boundary layers}.  \jt{J. Fluid Mech.}
  \bvol{30}~(02),  \pg{241--258}.

\bibitem[Climent \& Maxey(2009)]{climent2009force}
{\sc \au{Climent, E.} \& \au{Maxey, M.~R.}} \yr{2009} {\em The force coupling
  method: a flexible approach for the simulation of particulate flows\/}.
  \publ{inserted in `Theoretical Methods for Micro Scale Viscous Flows',
  Ressign Press, Eds F. Feuillebois and A. Sellier}.

\bibitem[Costa {\em et~al.\/}(2016)Costa, Picano, Brandt \&
  Breugem]{costa2016universal}
{\sc \au{Costa, P.}, \au{Picano, F.}, \au{Brandt, L.} \& \au{Breugem, W.-P.}}
  \yr{2016}  \at{Universal scaling laws for dense particle suspensions in
  turbulent wall-bounded flows}.  \jt{Phys. Rev. Lett.}  \bvol{117}~(13),
  \pg{134501}.

\bibitem[Del~{\'A}lamo \& Jimenez(2006)]{del2006linear}
{\sc \au{Del~{\'A}lamo, J.~C.} \& \au{Jimenez, J.}} \yr{2006}  \at{Linear
  energy amplification in turbulent channels}.  \jt{J. Fluid Mech.}
  \bvol{559},  \pg{205--213}.

\bibitem[Elghobashi \& Truesdell(1993)]{elghobashi1993two}
{\sc \au{Elghobashi, S.} \& \au{Truesdell, G.}} \yr{1993}  \at{On the two-way
  interaction between homogeneous turbulence and dispersed solid particles. i:
  Turbulence modification}.  \jt{Phys. Fluids A}  \bvol{5}~(7),
  \pg{1790--1801}.

\bibitem[Ellingsen \& Palm(1975)]{ellingsen1975stability}
{\sc \au{Ellingsen, T.} \& \au{Palm, E.}} \yr{1975}  \at{Stability of linear
  flow}.  \jt{Phys. Fluids}  \bvol{18}~(4),  \pg{487--488}.

\bibitem[Fornari {\em et~al.\/}(2016)Fornari, Formenti, Picano \&
  Brandt]{fornari2016effect}
{\sc \au{Fornari, W.}, \au{Formenti, A.}, \au{Picano, F.} \& \au{Brandt, L.}}
  \yr{2016}  \at{The effect of particle density in turbulent channel flow laden
  with finite size particles in semi-dilute conditions}.  \jt{Phys. Fluids}
  \bvol{28}~(3),  \pg{033301}.

\bibitem[Hamilton {\em et~al.\/}(1995)Hamilton, Kim \&
  Waleffe]{hamilton1995regeneration}
{\sc \au{Hamilton, J.~M.}, \au{Kim, J.} \& \au{Waleffe, F.}} \yr{1995}
  \at{Regeneration mechanisms of near-wall turbulence structures}.  \jt{J.
  Fluid Mech.}  \bvol{287},  \pg{317--348}.

\bibitem[Ho \& Leal(1974)]{ho1974inertial}
{\sc \au{Ho, B.} \& \au{Leal, L.}} \yr{1974}  \at{Inertial migration of rigid
  spheres in two-dimensional unidirectional flows}.  \jt{J. Fluid Mech.}
  \bvol{65}~(02),  \pg{365--400}.

\bibitem[Jim{\'e}nez(2011)]{jimenez2011cascades}
{\sc \au{Jim{\'e}nez, J.}} \yr{2011}  \at{Cascades in wall-bounded turbulence}.
   \jt{Annu. Rev. Fluid. Mech.}  \bvol{44}~(1),  \pg{27}.

\bibitem[Jim{\'e}nez \& Moin(1991)]{jimenez1991minimal}
{\sc \au{Jim{\'e}nez, J.} \& \au{Moin, P.}} \yr{1991}  \at{The minimal flow
  unit in near-wall turbulence}.  \jt{J. Fluid Mech.}  \bvol{225},
  \pg{213--240}.

\bibitem[Jim{\'e}nez \& Pinelli(1999)]{jimenez1999autonomous}
{\sc \au{Jim{\'e}nez, J.} \& \au{Pinelli, A.}} \yr{1999}  \at{The autonomous
  cycle of near-wall turbulence}.  \jt{J. Fluid Mech.}  \bvol{389},
  \pg{335--359}.

\bibitem[Klinkenberg {\em et~al.\/}(2013)Klinkenberg, Sardina, De~Lange \&
  Brandt]{klinkenberg2013numerical}
{\sc \au{Klinkenberg, J.}, \au{Sardina, G.}, \au{De~Lange, H.} \& \au{Brandt,
  L.}} \yr{2013}  \at{Numerical study of laminar-turbulent transition in
  particle-laden channel flow}.  \jt{Phys. Rev. E}  \bvol{87}~(4),
  \pg{043011}.

\bibitem[Lashgari {\em et~al.\/}(2015)Lashgari, Picano \&
  Brandt]{lashgari2015transition}
{\sc \au{Lashgari, I.}, \au{Picano, F.} \& \au{Brandt, L.}} \yr{2015}
  \at{Transition and self-sustained turbulence in dilute suspensions of
  finite-size particles}.  \jt{J. Theor. App. Mech. Pol.}  \bvol{5}~(3),
  \pg{121--125}.

\bibitem[Linares-Guerrero {\em et~al.\/}(2017)Linares-Guerrero, Hunt \&
  Zenit]{linares2017effects}
{\sc \au{Linares-Guerrero, E.}, \au{Hunt, M.~L.} \& \au{Zenit, R.}} \yr{2017}
  \at{Effects of inertia and turbulence on rheological measurements of
  neutrally buoyant suspensions}.  \jt{J. Fluid Mech.}  \bvol{811},
  \pg{525--543}.

\bibitem[Loisel {\em et~al.\/}(2013)Loisel, Abbas, Masbernat \&
  Climent]{loisel2013effect}
{\sc \au{Loisel, V.}, \au{Abbas, M.}, \au{Masbernat, O.} \& \au{Climent, E.}}
  \yr{2013}  \at{The effect of neutrally buoyant finite-size particles on
  channel flows in the laminar-turbulent transition regime}.  \jt{Phys. Fluids}
   \bvol{25}~(12),  \pg{123304}.

\bibitem[Majji {\em et~al.\/}(2016)Majji, Morris \& Banerjee]{majji2016flow}
{\sc \au{Majji, M.~V.}, \au{Morris, J.} \& \au{Banerjee, S.}} \yr{2016} Flow
  transition of neutrally buoyant suspension between concentric cylinders.
  \bt{In {\em ICTAM16, Montreal, Canada\/}}.

\bibitem[Matas {\em et~al.\/}(2003)Matas, Morris \&
  Guazzelli]{matas2003transition}
{\sc \au{Matas, J.-P.}, \au{Morris, J.~F.} \& \au{Guazzelli, E.}} \yr{2003}
  \at{Transition to turbulence in particulate pipe flow}.  \jt{Phys. Rev.
  Lett.}  \bvol{90},  \pg{014501}.

\bibitem[Orlandi \& Jim{\'e}nez(1994)]{orlandi1994generation}
{\sc \au{Orlandi, P.} \& \au{Jim{\'e}nez, J.}} \yr{1994}  \at{On the generation
  of turbulent wall friction}.  \jt{Phys. Fluids}  \bvol{6}~(2),
  \pg{634--641}.

\bibitem[Panton(2001)]{panton2001overview}
{\sc \au{Panton, R.~L.}} \yr{2001}  \at{Overview of the self-sustaining
  mechanisms of wall turbulence}.  \jt{Prog Aerosp Sci}  \bvol{37}~(4),
  \pg{341--383}.

\bibitem[Papavassiliou \& Hanratty(1997)]{papavassiliou1997interpretation}
{\sc \au{Papavassiliou, D.~V.} \& \au{Hanratty, T.~J.}} \yr{1997}
  \at{Interpretation of large-scale structures observed in a turbulent plane
  couette flow}.  \jt{Int. J. Heat Fluid Flow}  \bvol{18}~(1),  \pg{55--69}.

\bibitem[Picano {\em et~al.\/}(2015)Picano, Breugem \&
  Brandt]{picano2015turbulent}
{\sc \au{Picano, F.}, \au{Breugem, W.-P.} \& \au{Brandt, L.}} \yr{2015}
  \at{Turbulent channel flow of dense suspensions of neutrally buoyant
  spheres}.  \jt{J. Fluid Mech.}  \bvol{764},  \pg{463--487}.

\bibitem[Qureshi {\em et~al.\/}(2007)Qureshi, Bourgoin, Baudet, Cartellier \&
  Gagne]{qureshi2007turbulent}
{\sc \au{Qureshi, N.~M.}, \au{Bourgoin, M.}, \au{Baudet, C.}, \au{Cartellier,
  A.} \& \au{Gagne, Y.}} \yr{2007}  \at{Turbulent transport of material
  particles: an experimental study of finite size effects}.  \jt{Phys. Rev.
  Lett.}  \bvol{99}~(18),  \pg{184502}.

\bibitem[Schoppa \& Hussain(2002)]{schoppa2002coherent}
{\sc \au{Schoppa, W.} \& \au{Hussain, F.}} \yr{2002}  \at{Coherent structure
  generation in near-wall turbulence}.  \jt{J. Fluid Mech.}  \bvol{453},
  \pg{57--108}.

\bibitem[Shao {\em et~al.\/}(2012)Shao, Wu \& Yu]{shao2012fully}
{\sc \au{Shao, X.}, \au{Wu, T.} \& \au{Yu, Z.}} \yr{2012}  \at{Fully resolved
  numerical simulation of particle-laden turbulent flow in a horizontal channel
  at a low reynolds number}.  \jt{J. Fluid Mech.}  \bvol{693},  \pg{319--344}.

\bibitem[Stickel \& Powell(2005)]{stickel2005fluid}
{\sc \au{Stickel, J.~J.} \& \au{Powell, R.~L.}} \yr{2005}  \at{Fluid mechanics
  and rheology of dense suspensions}.  \jt{Annu. Rev. Fluid. Mech.}  \bvol{37},
   \pg{129--149}.

\bibitem[Townsend(1980)]{townsend1980structure}
{\sc \au{Townsend, A.~A.}} \yr{1980} {\em The structure of turbulent shear
  flow\/}.  \publ{Cambridge university press}.

\bibitem[Tuerke \& Jim{\'e}nez(2013)]{tuerke2013simulations}
{\sc \au{Tuerke, F.} \& \au{Jim{\'e}nez, J.}} \yr{2013}  \at{Simulations of
  turbulent channels with prescribed velocity profiles}.  \jt{J. Fluid Mech.}
  \bvol{723},  \pg{587--603}.

\bibitem[Waleffe(1997)]{waleffe1997self}
{\sc \au{Waleffe, F.}} \yr{1997}  \at{On a self-sustaining process in shear
  flows}.  \jt{Phys. Fluids}  \bvol{9}~(4),  \pg{883--900}.

\bibitem[Wallace(2016)]{wallace2016quadrant}
{\sc \au{Wallace, J.~M.}} \yr{2016}  \at{Quadrant analysis in turbulence
  research: History and evolution}.  \jt{Annu. Rev. Fluid. Mech.}  \bvol{48},
  \pg{131--158}.

\bibitem[Wang {\em et~al.\/}(2017)Wang, Abbas \& Climent]{wang2017modulation}
{\sc \au{Wang, G.}, \au{Abbas, M.} \& \au{Climent, E.}} \yr{2017}
  \at{Modulation of large-scale structures by neutrally buoyant and inertial
  finite-size particles in turbulent couette flow}.  \jt{Phys. Rev. Fluids}
  \bvol{2}~(8),  \pg{084302}.

\bibitem[Yu {\em et~al.\/}(2016)Yu, Vinkovic \& Buffat]{yu2016finite}
{\sc \au{Yu, W.}, \au{Vinkovic, I.} \& \au{Buffat, M.}} \yr{2016}
  \at{Finite-size particles in turbulent channel flow: quadrant analysis and
  acceleration statistics}.  \jt{J Turbul}  \bvol{17}~(11),  \pg{1048--1071}.

\bibitem[Yu {\em et~al.\/}(2013)Yu, Wu, Shao \& Lin]{yu2013numerical}
{\sc \au{Yu, Z.}, \au{Wu, T.}, \au{Shao, X.} \& \au{Lin, J.}} \yr{2013}
  \at{Numerical studies of the effects of large neutrally buoyant particles on
  the flow instability and transition to turbulence in pipe flow}.  \jt{Phys.
  Fluids}  \bvol{25}~(4),  \pg{043305}.

\end{thebibliography}


\end{document}